\documentclass[11pt, a4paper]{article}
\pdfoutput=1
\usepackage{booktabs}
\usepackage{subcaption}
\usepackage{jheppub}
\usepackage[T1]{fontenc}
\usepackage{float}
\usepackage{bm}
\usepackage{latexsym, amsmath, amstext, amssymb, amsfonts, amscd, bm, array, multirow, amsbsy, mathrsfs}
\usepackage{amsthm}
\usepackage{t1enc}
\usepackage{indentfirst}
\usepackage{pb-diagram}
\usepackage{graphicx}
\usepackage{enumerate}
\usepackage{color}
\usepackage[all]{xy}
\usepackage{hyperref}
\hypersetup{colorlinks=false,pdfborderstyle={/S/U/W 0}}

\usepackage{url}

\DeclareMathOperator{\arccot}{arccot}

\DeclareMathOperator{\arccosh}{arccosh}
\DeclareMathOperator{\arcsinh}{arcsinh}
\DeclareMathOperator{\arctanh}{arctanh}

\newcommand{\A}{{\Bbb{A}}}

\newcommand{\D}{\mathrm{D}}

\newcommand{\bl}{\begin{Lemma}}
	\newcommand{\el}{\end{Lemma}}
\newcommand{\bt}{\begin{Theorem}}
	\newcommand{\et}{\end{Theorem}}
\newcommand{\bd}{\begin{definition}}
	\newcommand{\ed}{\end{definition}}

\newcommand{\eqdef}{\stackrel{{\rm def.}}{=}}

\DeclareFontFamily{U}{rsf}{}
\DeclareFontShape{U}{rsf}{m}{n}{<5> <6> rsfs5 <7> <8> <9> rsfs7 <10-> rsfs10}{}
\DeclareMathAlphabet\Scr{U}{rsf}{m}{n}

\def\Z{\mathbb{Z}}
\def\C{\mathbb{C}}
\def\R{\mathbb{R}}

\def\H{\mathbb{H}}

\def\PSL{\mathrm{PSL}}

\def\dd{\mathrm{d}}

\def\vol{\mathrm{vol}}

\def\balpha{{\boldsymbol{\alpha}}}

\def\bt{\mathbf{t}}

\newcommand{\be}{\begin{equation*}}
\newcommand{\ee}{\end{equation*}}
\newcommand{\ben}{\begin{equation}}
\newcommand{\een}{\end{equation}}
\newcommand{\beqa}{\begin{eqnarray*}}
	\newcommand{\eeqa}{\end{eqnarray*}}
\newcommand{\beqan}{\begin{eqnarray}}
\newcommand{\eeqan}{\end{eqnarray}}

\newcommand{\nn}{\nonumber}

\newcommand{\Tr}{\mathrm{Tr}}

\def\cR{{\mathcal R}}

\def\cC{{\mathcal C}}

\def\cD{\mathcal{D}}

\def\hSigma{\widehat{\Sigma}}

\def\cE{\check{E}}

\def\cG{{\mathcal{G}}}

\def\cC{\mathcal{C}}

\def\G_2{\mathrm{G_2}}

\def\fC{\mathfrak{C}}

\def\fF{\mathfrak{F}}
\def\fI{\mathfrak{I}}
\def\fD{\mathfrak{D}}

\def\P{\mathbb{P}}

\def\mD{\mathbf{D}}

\def\G{\mathrm{G}}

\def\Im{\mathrm{Im}}

\def\cE{\mathcal{E}}

\def\hPhi{\widehat{\Phi}}

\def\mD{\mathbb{D}}
\def\fD{\mathfrak{D}}

\def\bd{\boldsymbol{\dd}}

\def\rS{\mathrm{S}}

\def\Re{\mathrm{Re}}
\def\Im{\mathrm{Im}}

\def\i{\mathbf{i}}
\def\sc{\mathrm{sc}}

\def\tPhi{{\widetilde \Phi}}
\def\tvarphi{{\widetilde \varphi}}

\def\fr{\frak{r}}
\def\ft{\frak{t}}

\def\rC{\mathrm{C}}
\def\rF{\mathrm{F}}
\def\rH{\mathrm{H}}
\def\tv{\widetilde{v}}
\def\balpha{\boldsymbol{\alpha}}

\def\btPhi{\boldsymbol{\tPhi}}

\title{Generalized $\alpha$-attractor models from elementary hyperbolic surfaces}

\author{Elena Mirela~Babalic, Calin Iuliu Lazaroiu}

\affiliation{Center for Geometry and Physics, Institute for
  Basic Science, Pohang 37673, Republic of Korea}

\emailAdd{mirela@ibs.re.kr, calin@ibs.re.kr}

\abstract{We consider generalized $\alpha$-attractor models whose
  scalar potentials are globally well-behaved and whose scalar
  manifolds are elementary hyperbolic surfaces. Beyond the Poincar\'e
  disk $\mD$, such surfaces include the hyperbolic punctured disk
  $\mD^\ast$ and the hyperbolic annuli $\A(R)$ of modulus $\mu=2\log
  R>0$. For each elementary surface, we discuss its decomposition into
  canonical end regions and give an explicit construction of the
  embedding into the Kerekjarto-Stoilow compactification (which in all three
  cases is the unit sphere), showing how this embedding allows for a
  universal treatment of globally well-behaved scalar potentials upon
  expanding their extension in real spherical harmonics. For certain
  simple but natural choices of extended potentials, we compute scalar
  field trajectories by projecting numerical solutions of the lifted
  equations of motion from the Poincar\'e half-plane through the
  uniformization map, thus illustrating the rich cosmological dynamics
  of such models.}

\begin{document}


\maketitle

\pagebreak


\section*{Introduction}
\label{Intro}

\noindent In \cite{genalpha}, we introduced a class of cosmological models which
form a wide generalization of $\alpha$-attractors
\cite{alpha1,alpha2,alpha3,alpha4,alpha5,Escher}. Such models are
defined by cosmological solutions (with flat spatial section) of the
Einstein theory of gravity coupled to two real scalar fields, the
dynamics of the latter being described by a non-linear sigma model
whose scalar manifold is a non-compact, oriented and topologically
finite borderless surface $\Sigma$ endowed with a complete metric
$\cG$ of constant Gaussian curvature $K=-\frac{1}{3\alpha}$. The
scalar potential is given by a smooth real-valued function $\Phi$
defined on $\Sigma$. The rescaled scalar field metric $G\eqdef
\frac{1}{3\alpha}\cG$ has Gaussian curvature equal to $-1$, hence
$(\Sigma,G)$ is a geometrically finite hyperbolic surface in the sense
of \cite{Borthwick}. Using the uniformization theorem of Koebe and
Poincar\'e \cite{unif} and the theory of surface groups (Fuchsian
groups without elliptic elements), reference \cite{genalpha} discussed
the basic features of cosmological dynamics and inflation in such models.

In the present paper, we consider the non-generic situation when
$(\Sigma,G)$ is elementary, focusing on special features which arise
in this case. As explained in \cite{genalpha}, an elementary
hyperbolic surface $(\Sigma,G)$ is isometric with the hyperbolic disk
$\mD$, with the hyperbolic punctured disk $\mD^\ast$ or with the
hyperbolic annulus $\A(R)$ of modulus $\mu=2\log R$, where $R>1$ is a
real number. Each of these surfaces is planar, having the unit sphere
$\rS^2$ as its end compactification. This allows one to parameterize
globally well-behaved \cite{genalpha} scalar potentials $\Phi$ through
the coefficients of the Laplace expansion of their global extension
$\hPhi$ to the end compactification, which in this case reduces to an
expansion in real spherical harmonics. Both the hyperbolic metric and
a fundamental polygon are known explicitly for $\mD^\ast$ and
$\A(R)$. In particular, one can describe explicitly the hyperbolic
geometry of such surfaces and one can compute scalar field
trajectories on $(\Sigma,\cG)$ by determining trajectories of an
appropriate lift of the model to the Poincar\'e half-plane $\H$ and
projecting them to $\mD^\ast$ or to $\A(R)$ through the uniformization
map.  When the scalar potential is constant, this illustrates how the
different projections from $\H$ to $\mD^\ast$ and $\A(R)$ can take the
same trajectory on the Poincar\'e half-plane into qualitatively
different trajectories on the corresponding elementary hyperbolic
surface. The explicit embeddings of $\mD$, $\mD^\ast$ and $\A(R)$ into
their common Kerekjarto-Stoilow compactification $\rS^2$ are also
different. As a consequence, a smooth real-valued function defined on
$\rS^2$ will generally induce rather different globally well-behaved
scalar potentials on the disk, the punctured disk and the annulus.
When combined with the different projections from $\H$, this leads to
qualitatively different dynamics on $\mD$, $\mD^\ast$ and $\A(R)$. By
considering some simple but natural choices of extended scalar
potentials on $\rS^2$, we use numerical methods to compute various
trajectories on $\mD^\ast$ and $\A(R)$, thus illustrating the rich
dynamics of such models, which is not entirely visible in the gradient
flow approximation.

The paper is organized as follows. Section 1 briefly recalls the
definition of generalized $\alpha$-attractor models and the lift of
their cosmological evolution equation to the Poincar\'e half-plane. In
Section 2, we consider globally well-behaved scalar potentials on
topologically-finite planar oriented surfaces (of which the elementary
surfaces are particular cases), showing how they can be parameterized
through the coefficients of the Laplace expansion of their extension
to the end compactification. In Section 3, we review the
classification of elementary hyperbolic surfaces and some of their
basic properties. Section 4 recalls the case of the hyperbolic disk,
showing how it fits into the general approach developed in
\cite{genalpha} and how well-behaved scalar potentials can be
described through the Laplace expansion of their extension to
$\rS^2$. Section 5 discusses the hyperbolic punctured disk. After
giving the explicit form of the hyperbolic metric on $\mD^\ast$, of a
partial isometric embedding into three-dimensional Euclidean space and
of the decomposition into horn and cusp regions, we compute certain
scalar field trajectories for a few globally well-behaved scalar
potentials which are induced naturally from $\rS^2$. This illustrates
the rich dynamics of our models on the punctured hyperbolic disk. We
also discuss inflationary regions and the number of e-folds for
various trajectories and provide an explicit example of an
inflationary trajectory which produces between 50 and 60 e-folds, thus
showing that such models can match observational constraints.
 Section 6 performs a similar analysis for hyperbolic
annuli, using globally well-behaved scalar potentials induced by the
same choices of functions on $\rS^2$. In particular, this illustrates
how the dynamics of our models differs on $\A(R)$ and
$\mD^\ast$. Section 7 comments briefly on the relation of such models
with observational cosmology, while Section 8 contains our conclusions
and sketches a few directions for further research.

\paragraph{Notations and conventions.} 
All manifolds considered are smooth, connected, oriented and
paracompact (hence also second-countable). All homeomorphisms and
diffeomorphisms considered are orientation-preserving. By definition,
a Lorentzian four-manifold has ``mostly plus'' signature. The symbol
$\i$ denotes the imaginary unit. The Poincar\'e half-plane is 
the upper half plane with complex coordinate $\tau$: 
\ben
\label{PoincareHP}
\H=\{\tau\in \C\,|\,\Im\tau>0\}~~,
\een
endowed with its unique complete metric of Gaussian curvature $-1$,
which is given by:
\be
\dd s^2_\H=\frac{1}{(\Im \tau)^2}|\dd \tau|^2~~.
\ee
The real coordinates on $\H$ are denoted by $x\eqdef \Re \tau$ and
$y\eqdef \Im \tau$. The complex coordinate on the hyperbolic disk, 
the hyperbolic punctured disk or an annulus is denoted by $u$.
We define the {\em rescaled Planck mass} through:
\ben
\label{M0}
M_0\eqdef M\sqrt{\frac{2}{3}}~~,
\een
where $M$ is the (reduced) Planck mass. 

\section{Generalized $\alpha$-attractor models}

\subsection{Definition of the models}

\noindent Let $(\Sigma, G)$ be a non-compact oriented, connected and
complete two-dimensional Riemannian manifold without boundary (called
the {\em scalar manifold}) and $\Phi:\Sigma\rightarrow \R$ be a smooth
function (called the {\em scalar potential}). We say that $(\Sigma,G)$
is {\em hyperbolic} if the metric $G$ has constant Gaussian curvature
equal to $-1$. We say that $\Sigma$ is {\em topologically finite} and that
$(\Sigma,G)$ is {\em geometrically finite} if the fundamental group
$\pi_1(\Sigma)$ is finitely-generated. Let $\cG=3\alpha G$, where $\alpha>0$ 
is a fixed positive real number.

Let $X$ be any four-manifold which can support Lorentzian metrics. The
Einstein-Scalar theory defined by the triplet $(\Sigma,\cG,\Phi)$ includes
four-dimensional gravity (described by a Lorentzian metric $g$ defined
on $X$) and a smooth map $\varphi:X\rightarrow \Sigma$, with action
\cite{genalpha}:
\ben
\label{S}
S[g,\varphi]=\int_X \left[\frac{M^2}{2} \mathrm{R}(g)-\frac{1}{2}\Tr_g \varphi^\ast(\cG)-\Phi\circ \varphi\right] \vol_g~~.
\een
Here $\vol_g$ is the volume form of $(X,g)$, $\mathrm{R}(g)$ is the
scalar curvature of $g$ and $M$ is the (reduced) Planck mass.

When $X$ is diffeomorphic with $\R^4$ and $g$ is a FLRW metric with
flat spatial section, solutions of the equations of motion for the
action \eqref{S} for which $\varphi$ depends only on the cosmological
time $t$ define the class of so-called {\em generalized
  $\alpha$-attractor models} \cite{genalpha}. In this case, the map
$\fI\ni t\rightarrow \varphi(t)\in \Sigma$ (where $\fI$ is a real
interval) defines a curve in $\Sigma$ which obeys an
invariantly-defined non-linear second order ordinary differential
equation (which is locally equivalent with a system of two second order
equations). We refer the reader to \cite{genalpha} for a
general discussion of such models.

\subsection{Lift to the Poincar\'e half-plane}
\label{subsec:lift}

\noindent As explained in \cite{genalpha}, the cosmological equations of motion
can be lifted from $\Sigma$ to the Poincar\'e half-plane $\H$ by using
the covering map $\pi_\H:\H\rightarrow \Sigma$ which uniformizes
$(\Sigma,G)$ to $\H$.  This allows one to determine the cosmological
trajectories $\varphi(t)$ by projecting to $\Sigma$ the trajectories
$\tvarphi(t)$ of a ``lifted'' model defined on $\H$. The lifted model
is governed by the following system of second order non-linear
ordinary differential equations \cite[eq. (7.4)]{genalpha}:
\beqan
\label{elplane}
&& \ddot{x}-\frac{2}{y}\dot{x}\dot{y} +\frac{1}{M} \sqrt{\frac{3}{2}} 
\left[3\alpha \frac{\dot{x}^2+\dot{y}^2}{y^2}+2\tPhi(x,y)\right]^{1/2}\!\!\dot{x}+\frac{1}{3\alpha} y^2 \partial_x\tPhi(x,y)=0 \\
&& \ddot{y}+\frac{1}{y}(\dot{x}^2-\dot{y}^2)+\frac{1}{M} \sqrt{\frac{3}{2}} \left[3\alpha \frac{\dot{x}^2+\dot{y}^2}{y^2}+2\tPhi(x,y)\right]^{1/2}\!\!\dot{y}+\frac{1}{3\alpha} y^2\partial_y\tPhi(x,y)=0~~,\nn
\eeqan
where $\dot{~}\eqdef \frac{\dd}{\dd t}$\,, $t$ is the cosmological time 
while $x=\Re\tau$, $y=\Im \tau$
are the Cartesian coordinates on the Poincar\'e half plane with
complex coordinate $\tau$ and $\tPhi\eqdef \Phi\circ \pi_\H:\H\rightarrow \R$ 
is the {\em lifted potential}. Let $u_0$ be any point of $\Sigma$ and let
$\tau_0\in \H$ be chosen such that $\pi_\H(\tau_0)=u_0$. An initial
velocity vector $v_0=\dot{u}_0 \in T_{u_0}\Sigma$ defined at $u_0$
on $\mD^\ast$ and its unique lift $\tv_0=\dot{\tau}_0\in
T_{\tau_0}\H$ through the differential\footnote{Notice that the
  differential of $\pi_\H$ at $\tau_0$ is a bijective linear map
  because $\pi_\H$ is a covering map and hence a local
  diffeomorphism.} of $\pi_\H$ at $\tau_0$ are related through:
\be
v_0=(\dd_{\tau_0}\pi_\H)(\tv_0)~~.
\ee
Writing $\tau=x+\i y$ and $\tau_0=x_0+\i y_0$, we have
$\tv_0=\tv_{0x}+\i \tv_{0y}$ with real $\tv_{0x}$, $\tv_{0y}$.  As
shown in \cite{genalpha}, a cosmological trajectory $\varphi(t)$ on
$\Sigma$ with initial condition $(u_0,\tau_0)$ can be written as
$\varphi(t)=\pi_\H(\tvarphi(t))$, where $\tvarphi(t)=x(t)+\i y(t)$ is
the
solution of the lifted system \eqref{elplane} with initial conditions:
\be
x(0)=x_0~~,~~y(0)=y_0~~\mathrm{and}~~\dot{x}(0)=\tv_{0x}~~,~~\dot{y}(0)=\tv_{0y}~~.
\ee

\paragraph{Eliminating the Planck mass.}

Let $\btPhi=\frac{1}{M_0}\tPhi$ and $\balpha=\frac{1}{M_0}\alpha$,
where $M_0=M\sqrt{\frac{2}{3}}$ is the rescaled Planck mass
\eqref{M0}. Then \eqref{elplane} becomes:
\beqan
\label{elplane0}
&& \ddot{x}-\frac{2}{y}\dot{x}\dot{y} +\left[3\balpha \frac{\dot{x}^2+\dot{y}^2}{y^2}+2\btPhi(x,y)\right]^{1/2}
\!\!\dot{x}+\frac{1}{3\balpha} y^2 \partial_x\btPhi(x,y)=0~~, \nn\\
&& \ddot{y}+\frac{1}{y}(\dot{x}^2-\dot{y}^2)+\left[3\balpha \frac{\dot{x}^2+\dot{y}^2}{y^2}+2\btPhi(x,y)\right]^{1/2}\!\!\dot{y}+\frac{1}{3\balpha} y^2\partial_y\btPhi(x,y)=0~~,
\eeqan
showing that we can eliminate the Planck mass from the
equations provided that we measure both $\alpha$ and $\tPhi$ (and
hence also $\Phi$) in units of $M_0$. The numerical solutions extracted 
in latter sections of this paper were obtained using the system 
\eqref{elplane0}, after performing such a rescaling of $\alpha$ and 
$\Phi$.

\section{Laplace expansion of globally well-behaved scalar potentials}
\label{sec:Laplace}

\noindent Let $\hSigma$ denote the Kerekjarto-Stoilow (a.k.a. end)
compactification of $\Sigma$ (see \cite{Richards, Stoilow}) and
identify $\Sigma$ with its image in $\hSigma$ through the embedding map
$j:\Sigma\rightarrow \hSigma$.  A smooth scalar potential
$\Phi:\Sigma\rightarrow \R$ is called {\em globally well-behaved}
\cite{genalpha} if there exists a smooth function
$\hPhi:\hSigma\rightarrow \R$ whose restriction to $\Sigma$ equals
$\Phi$, in which case $\hPhi$ is uniquely determined by $\Phi$ through
continuity. Since all elementary hyperbolic surfaces $(\Sigma,G)$ are
planar, their end compactification $\hSigma$ is diffeomorphic with the
unit sphere $\rS^2=\{(x_1,x_2,x_3)\in \R^3 \, | \,
x_1^2+x_2^2+x_3^2=1\}$, so in this case globally well-defined scalar
potentials on $\Sigma$ correspond bijectively to smooth functions
$\hPhi:\rS^2\rightarrow \R$\,.

Let $\psi$ and $\theta$ be spherical coordinates on $\rS^2$, thus:
\ben
\label{sphcoord}
x_1=\sin\psi\cos\theta~~,~~x_2=\sin\psi\sin\theta~~,~~x_3=\cos\psi~~,
\een
where $\psi\in [0,\pi]$ and $\theta\in [0,2\pi)$. Any smooth map
  $\hPhi:\rS^2\rightarrow \R$ is square-integrable with respect to the
  round Lebesgue measure $\sin\psi \dd\psi\dd\theta$ on $\rS^2$ and
  admits the Laplace-Fourier series expansion:
\ben
\label{sharm}
\hPhi(\psi,\theta)=\sum_{l=0}^\infty \sum_{m=-l}^l C_{lm} Y_l^m(\psi,\theta)~~, 
\een
where $Y_l^m$ are the complex spherical harmonics and: 
\be
C_{lm}= \int_{\rS^2} \dd\psi\dd\theta \sin\psi \overline{Y_l^m(\psi,\theta)} \hPhi(\psi,\theta)\in \C~~.
\ee
The series in \eqref{sharm} converges {\em uniformly}
to $\hPhi$ on $\rS^2$ since $\hPhi$ is smooth (see
\cite{Kalf}). Recall that $Y_l^m(\psi,\theta)=P_l^m(\cos\psi) e^{\i m
  \theta}$, where $P_l^m(x)$ are the associated Legendre
functions. Since $\hPhi$ is real-valued, expansion \eqref{sharm}
reduces to:
\be
\hPhi(\psi,\theta)=\sum_{l=0}^\infty 
\sum_{m=-l}^l P_l^m(\cos\psi)[A_{lm}\cos(m\theta)+B_{lm}\sin(m\theta)]~~,
\ee
where $A_{lm}$ and $B_{lm}$ are real constants. Equivalently, we have: 
\ben
\label{hPhiExp}
\hPhi(\psi,\theta)=\sum_{l=0}^\infty \sum_{m=-l}^l D_{lm} Y_{lm}(\psi,\theta)~~,
\een
where $D_{lm}\in\R$ and $Y_{lm}$ are the real (a.k.a. tesseral) spherical
harmonics, which correspond to orbitals. This expansion is again
uniformly-convergent and gives a systematic way to approximate
$\hPhi$ by truncating away the contributions with $l$ greater than
some cutoff value.

\paragraph{Some particular choices for $\hPhi$.}

If only the modes with $l=0$ and $l=1$ ($s$ and $p$
orbitals) are present, then we have:
\ben
\label{hPhiOrb}
\hPhi(\psi,\theta)=a+b\cos\psi+c\sin\psi \cos\theta+d\sin\psi\sin\theta~~,
\een
where $a,b,c,d$ are real constants and we used the expressions: 
\beqa
&& P_0^0(x)=1~~,~~P_1^0(x)=x~~,~~P_1^1(x)=-\sqrt{1-x^2}~~,~~P_1^{-1}(x)=\frac{1}{2}\sqrt{1-x^2}~\\
&& Y_{00}=\frac{1}{2\sqrt{\pi}}~~,~~Y_{10}=\sqrt{\frac{3}{4\pi}}\cos \psi~~,~~Y_{11}=\sqrt{\frac{3}{4\pi}}\sin \psi\cos\theta~~,~~Y_{1,-1}=\sqrt{\frac{3}{4\pi}}\sin \psi\sin\theta~~.
\eeqa
The constant term in \eqref{hPhiOrb} corresponds to the $s$ orbital
$(Y_{00}$) while the terms with prefactors $b$, $c$ and $d$ correspond
to the orbitals $p_z$ ($Y_{10}$), $p_x$ ($Y_{11}$) and $p_y$
($Y_{1,-1}$).

\begin{figure}[H]
\centering \includegraphics[width=100mm]{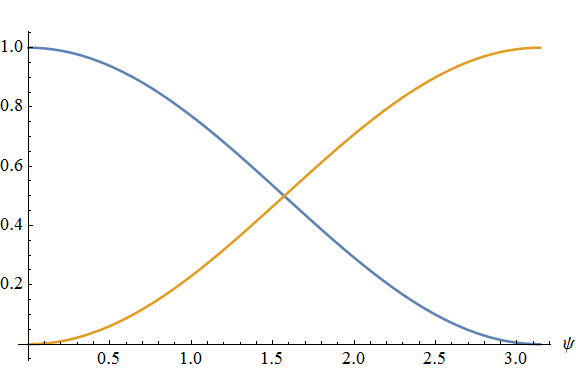}
\caption{Plot of $\hPhi_+/M_0$ (blue) and $\hPhi_-/M_0$ (yellow) as 
  functions of $\psi\in [0,\pi]$. The value $\psi=0$
  corresponds to the north pole of $\rS^2$, while $\psi=\pi$
  corresponds to the south pole.}
\label{fig:hPhipm}
\end{figure}

For $c=d=0$, two simple choices are $a=b=\frac{1}{2} M_0$ and
$a=-b=\frac{1}{2}M_0$, where $M_0$ is the rescaled Planck mass
\eqref{M0}.  These give the following $\theta$-independent potentials,
which involve only the orbitals $s$ and $p_z$ and are shown in Figure
\ref{fig:hPhipm}:
\ben
\label{hPhipm}
\hPhi_+(\psi)=M_0\cos^2\left(\frac{\psi}{2}\right)~~,~~\hPhi_-(\psi)=M_0\sin^2\left(\frac{\psi}{2}\right)~~.
\een
Notice that $\hPhi_+$ has a maximum at $\psi=0$ (north pole) and a
minimum at $\psi=\pi$ (south pole) while $\hPhi_-$ has a minimum at
$\psi=0$ (north pole) and a maximum at $\psi=\pi$ (south pole).

Another simple choice is $b=d=0$ and $a=c=M_0$, which corresponds to
a linear combination of the $s$ and $p_x$ orbitals and gives the
extended potential:
\ben
\label{hPhi0}
\hPhi_0(\psi,\theta)=M_0(1+\sin\psi\cos\theta)=M_0(1+x_1)~~.
\een
Unlike $\hPhi_\pm$, this function does not have extrema at the north
or south pole.  Instead, it has two extrema along the equator of
$\rS^2$, having a maximum (equal to $2M_0$) at the point
$(\psi,\theta)=(\frac{\pi}{2},0)$ and a minimum (equal to $0$) at
$(\psi,\theta)=(\frac{\pi}{2},\pi)$. Notice that $\hPhi_\pm$ and
$\hPhi_0$ are Morse functions on $\rS^2$, so the potentials derived
from them on a planar surface (whose end compactification is $\rS^2$)
will be compactly Morse in the sense of \cite{genalpha}.

\paragraph{A universal approach to globally well-behaved potentials.}

The techniques of passing to the end compactification and lifting to
the Poincar\'e half-plane introduced in \cite{genalpha} allow for a
uniform treatment of globally well-behaved scalar potentials in
generalized $\alpha$-attractor models for which $(\Sigma,G)$ is
geometrically finite. This is summarized in the commutative diagram
\eqref{diagram:pots}, where $j$ is the smooth embedding of $\Sigma$
into its end compactification $\hSigma$ and $\pi_\H$ is the
uniformizing map from $\H$.
\ben
\label{diagram:pots}
\scalebox{1.0}{
\xymatrix{
\H~\ar_{\tPhi}[rd] \ar^{\pi_\H}[r]~ &~\Sigma ~\ar[d]^\Phi \ar[r]^{j} &\hSigma \ar^{\hPhi}[ld]\\
& \R & 
  }}
\een
Any smooth real-valued function $\hPhi$ defined on $\hSigma$ induces a
globally well-behaved scalar potential $\Phi$ on $\Sigma$ through the
formula $\Phi=\hPhi\circ j$, while any globally well-behaved scalar
potential $\Phi$ on $\Sigma$ lifts to a smooth function $\tPhi=\Phi\circ
\pi_\H$ defined on $\H$. When the maps $j$ and $\pi_\H$ are known,
$\tPhi$ can be recovered from $\hPhi$ as the composition
$\tPhi=\hPhi\circ \chi$, where $\chi:\H\rightarrow \hSigma$ is the composite
map $\chi\eqdef j\circ \pi_\H$. Notice that $j$ and $\chi$ are smooth 
maps, while $\pi_\H$ is holomorphic when $\Sigma$ is endowed with the complex 
structure which corresponds to the conformal class of the metric $G$. 
The maps $j,\pi_\H$ and $\chi$ differ for
distinct geometrically-finite hyperbolic surfaces $(\hSigma,G)$ having
the same end compactification $\hSigma$, which means that the same
``universal'' extended potential $\hPhi$ defined on $\hSigma$ can
induce markedly different globally well-behaved potentials $\Phi$ and
lifted potentials $\tPhi$ for different hyperbolic surfaces of the
same genus. 

When $\Sigma$ is a planar surface, the end compactification $\hSigma$
coincides with the unit sphere and $\hPhi$ can be expanded into real
spherical harmonics as explained above. This induces
uniformly-convergent expansions:
\beqa
&&\Phi=\sum_{l=0}^\infty \sum_{m=-l}^l D_{lm} Y_{lm}\circ j\\
&&\tPhi=\sum_{l=0}^\infty \sum_{m=-l}^l D_{lm} Y_{lm}\circ \chi
\eeqa
of $\Phi$ and $\tPhi$. In the next sections, we determine explicitly
the maps $j$, $\pi_\H$ and $\chi$ for the planar elementary surfaces
$\mD$, $\mD^\ast$ and $\A(R)$ and the maps $\Phi$ and $\tPhi$ induced
by the choices $\hPhi=\hPhi_\pm$ and $\hPhi=\hPhi_0$ given above. This
illustrates how the same function $\hPhi$ leads to different dynamics
of the generalized $\alpha$-attractor models associated to distinct
planar surfaces. For the three elementary surfaces, we will construct the map
$j:\Sigma\rightarrow \hSigma=\rS^2$ by first diffeomorphically (but
not bi-holomorphically !) identifying $\Sigma$ with the complex plane $\C$
of complex coordinate $\zeta$ (or with $\C$ with a point removed) and
then identifying the latter with the once- or twice-punctured sphere by
using stereographic projection from the north pole of $\rS^2$:
\ben
\label{sterproj}
\zeta=\cot\left(\frac{\psi}{2}\right)e^{\i\theta} \in \C~~.
\een

\section{Elementary hyperbolic surfaces}
\label{sec:elem}

\noindent A (complete) hyperbolic surface $(\Sigma, G)$ is called {\em elementary} if it
is conformally-equivalent with a simply-connected or doubly-connected\footnote{A 
regular  domain $\cD\subset \C$ is called {\em doubly-connected} if its
  complement in the Riemann sphere has two connected components, which
  happens iff $\pi_1(\cD)\simeq \Z$.}
regular domain contained in the complex plane. This amounts to the condition
that the uniformizing surface group $\Gamma$ of $(\Sigma, G)$ is the
trivial group or a cyclic subgroup of $\PSL(2,\R)$ of parabolic or
hyperbolic type.

Any simply connected regular domain is conformally equivalent with the
unit disk $\mD=\{u\in \C\, |\, |u|<1\}$ (and hence with the upper half
plane $\H$). Such a domain admits a unique complete hyperbolic metric,
known as the Poincar\'e metric. Any doubly-connected regular domain is
conformally equivalent to one of the following, when the latter 
is endowed with the complex structure inherited from the complex plane:
\begin{itemize}
\itemsep 0.0em
\item The punctured plane $\C^\ast=\C\setminus\{0\}$
\item The punctured unit disk $\mD^\ast\eqdef \mD\setminus \{0\}$
\item The annulus $\A(R)\eqdef\{u\in \C\,\, |\,\, \frac{1}{R}<|u|<
  R\}$ of modulus $\mu=2\log R>0$, where $R>1$.
\end{itemize}
When endowed with its usual complex structure, the punctured plane
does not support a complete hyperbolic metric. On the other hand, the
punctured disk $\mD^\ast$ and annulus $\A(R)$ admit
uniquely-determined complete hyperbolic metrics. Notice that both
$\mD^\ast$ and $\A(R)$ are homeomorphic with (open) cylinders. Due to
this fact, the hyperbolic punctured disk $\mD^\ast$ is also called the
{\em parabolic cylinder} while the hyperbolic annuli $\A(R)$ are also
called {\em hyperbolic cylinders} \cite{Borthwick}. Summarizing, 
the list of elementary hyperbolic surfaces is as follows:
\begin{enumerate}
\itemsep 0.0em
\item The hyperbolic disk $\mD$ (which is isometric with the
  Poincar\'e half-plane $\H$).
\item The hyperbolic punctured disk $\mD^\ast$ (uniformized to $\H$ by
  a parabolic cyclic subgroup of $\PSL(2,\R)$).
\item The hyperbolic annuli $\A(R)$ for $R>1$ (uniformized to $\H$ by a
  hyperbolic cyclic subgroup of $\PSL(2,\R)$).
\end{enumerate}
The explicit form of the hyperbolic metric is known in all cases, as
is a fundamental polygon \cite{Beardon, Katok, FrenchelNielsen} for
$\mD^\ast$ and $\A(R)$. This allows one to study the cosmological
dynamics of $\alpha$-attractor models defined by such surfaces either
directly on $(\Sigma,G)$ or (as explained in \cite{genalpha}) by
lifting to the hyperbolic disk or to the Poincar\'e half-plane.

For elementary hyperbolic surfaces, the isometry classification 
of the ends is as follows \cite{genalpha,
  Borthwick, Drawing}:
\begin{itemize}
\itemsep 0.0em
\item The hyperbolic disk $\mD$ has a single end, known as a {\em
  plane end}.
\item The hyperbolic punctured disk $\mD^\ast$ has two ends. One of
  these is a {\em cusp end}, the other being a {\em horn end}.
\item The hyperbolic annulus $\A(R)$ has two ends, both of which are
  {\em funnel ends}.
\end{itemize}

All elementary hyperbolic surfaces are planar (i.e. of genus zero). As
explained in \cite{genalpha}, this implies that their end
compactification \cite{Richards, Stoilow} is the unit sphere
$\rS^2$. On the other hand, the conformal boundary $\partial_\infty^G
\Sigma$ \cite{genalpha, Maskit, Haas} differs in the three cases:
\begin{itemize}
\itemsep 0.0em
\item For the hyperbolic disk, we have $\partial_\infty \mD=\rS^1$,
  where the circle corresponds to the plane end.
\item For the hyperbolic punctured disk, we have $\partial_\infty
  \mD^\ast=\{0\}\sqcup \rS^1$, where the origin corresponds to the cusp end
  and $\rS^1$ corresponds to the horn end.
\item For the hyperbolic annulus, we have $\partial_\infty
  \A(\R)=\rS^1\sqcup \rS^1$, each of the circles corresponding to a
  funnel end.
\end{itemize}

\paragraph{Remark.}
By a theorem of Hilbert, a complete hyperbolic surface cannot be
embedded isometrically into Euclidean 3-space. However, incomplete
regions of such a surface can be embedded isometrically (and we
shall see examples of such partial embeddings in latter
sections). Notice that one can sometimes find isometric embeddings of
complete hyperbolic surfaces into {\em non-Euclidean} three-space,
such as the well-known embedding of the Poincar\'e disk as a sheet of a
hyperboloid defined inside three-dimensional Minkowski space.

\section{The hyperbolic disk}

\noindent The cosmological model defined by the hyperbolic disk $\mD$ coincides
with the two-field $\alpha$-attractor model of \cite{Escher}, which
was discussed extensively in the literature. The purpose of this
section is to show how this fits into the general theory developed in
\cite{genalpha}.

\subsection{Semi-geodesic coordinates}

\noindent
The unit disk $\mD=\{u\in \C \, | \, |u|<1\}$ admits a unique complete
hyperbolic metric, which is given by:
\ben
\label{gD}
\dd s_\mD^2=\lambda^2_\mD(u,\bar{u})|\dd u|^2~~,~\mathrm{where}~~\lambda_\mD(u,\bar{u})=\frac{2}{1-|u|^2}~~.
\een
In polar coordinates given by $\rho=|u|\in (0,1)$ and
$\theta=\arg(u)\in (0,2\pi)$, the metric becomes: 
\be
\dd s_\mD^2=\frac{4}{(1-\rho^2)^2}\left(\dd \rho^2+\rho^2\dd \theta^2\right)~~.
\ee
Semi-geodesic coordinates $(r,\theta)$ for $\mD$ are obtained by the
change of variables:
\be
\rho\eqdef \tanh\left(\frac{r}{2}\right)\in (0,1)~~\mathrm{i.e.}~~r=2\arctanh(\rho)=\log\frac{1+\rho}{1-\rho}\in (0,+\infty)~~.
\ee
This maps the unit disk (diffeomorphically, but not conformally) to
the complex plane with polar coordinates $(r,\theta)$ and complex
coordinate: 
\be
\zeta=re^{\i\theta}=2\arctanh(\rho)e^{\i\theta}
\ee
and brings the metric to the form:
\ben
\label{pmetric}
\dd s_\mD^2=\dd r^2+\sinh^2(r)\dd\theta^2~~.
\een
The single end of $\mD$ (which is called a {\em plane end})
corresponds to $r\rightarrow +\infty$, while the center of $\mD$
corresponds to $r\rightarrow 0$.

\subsection{The end compactification of $\mD$}

\noindent
The end compactification $\widehat{\mD}$ of the hyperbolic
disk coincides with the Alexandroff compactification of the
$\zeta$-plane, which by the stereographic projection \eqref{sterproj} 
is identified with the unit sphere $\rS^2$. The north pole
$\psi=0$ corresponds to the plane end at $r=|\zeta|\rightarrow
\infty$, while the south pole $\psi=\pi$ corresponds to $r=0$, i.e. to
the center of $\mD$. In spherical stereographic coordinates
$(\psi,\theta)$, the Poincar\'e metric \eqref{pmetric} becomes:
\be
\dd s_\mD^2=\frac{1}{4}\frac{\dd \psi^2}{\sin^2(\frac{\psi}{2})}+\sinh^2(\cot(\frac{\psi}{2}))\dd \theta^2~~.
\ee

\paragraph{Remark.}
The compact Riemann surface into which $\mD=\{u\in \C \, | \, |u|<1\}$
embeds holomorphically \cite{Maskit, Haas} is the Riemann sphere
$\C\P^1$ associated to the $u$-plane. The coordinate $u$ is given by:
\be
u\eqdef \frac{u_1}{u_2}~~\mathrm{for}~~u_2\neq 0~~,
\ee
where $(u_1,u_2)\in \C^2\setminus \{0,0\}$ are the homogeneous
coordinates of $\C\P^1$.

\subsection{Globally well-behaved scalar potentials on $\mD$}

\noindent
A potential $\Phi:\mD\rightarrow \R$ is globally well-behaved on $\mD$
iff there exists a smooth function $\hPhi:\rS^2\rightarrow \R$ such
that:
\ben
\label{PhibPhi}
\Phi(r,\theta)=\hPhi(2\arccot(r),\theta)~~,
\een
i.e.: 
\be
\Phi(\rho,\theta)=\hPhi(2\arccot(2\arctanh\rho),\theta)~~.
\ee
The condition that $\hPhi$ is smooth on $\rS^2$ implies, in
particular, that $\hPhi(\psi,\theta)$ has a finite limit for
$\psi\rightarrow 0$ and hence $\Phi(r,\theta)$ has a
$\theta$-independent limit for $r\rightarrow +\infty$ i.e. for
$|u|\rightarrow 1$. Expansion \eqref{hPhiExp} gives the
uniformly-convergent series:
\be
\Phi(\rho,\theta)=\sum_{l=0}^\infty \sum_{m=-l}^l D_{lm} Y_{lm}(2\arccot(2\arctanh\rho),\theta)~~.
\ee
To obtain inflationary behavior with the scalar field rolling from the
plane end towards the interior of $\mD$, one can require that $\hPhi$
has a local maximum at the north pole of $\rS^2$. In the simplest
models, one can take $\hPhi$ to have only two critical points, namely
a global maximum at the north pole and a global minimum at the south
pole. In that case, $\Phi$ has a global minimum at $u=0$ (the center
of $\mD$) and increases monotonically to a $\theta$-independent finite
value as $|u|$ grows from zero to $1$ (toward the conformal boundary
$\partial_\infty\mD=\rS^1$ of $\mD$).

\begin{figure}[H]
\centering
\begin{minipage}{.47\textwidth}
\centering \includegraphics[width=1\linewidth]{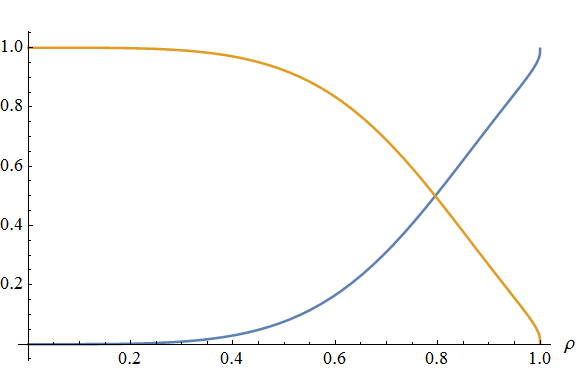}
\vskip 0.45em \subcaption{Plot of $\Phi_+/M_0$ (blue) and $\Phi_-/M_0$
  (yellow) as functions of $\rho\in [0,1)$. The value $\rho=0$
    corresponds to the center of $\mD$, while $\rho=1$ corresponds to
    the conformal boundary $\partial_\infty\mD$. Only the potential
    $\Phi_+$ leads to $\alpha$-attractor behavior when inflation takes
    place near the plane end.}
\label{fig:PhipmDisk}
\end{minipage}\hfill
\begin{minipage}{.47\textwidth}
\centering \includegraphics[width=0.95\linewidth]{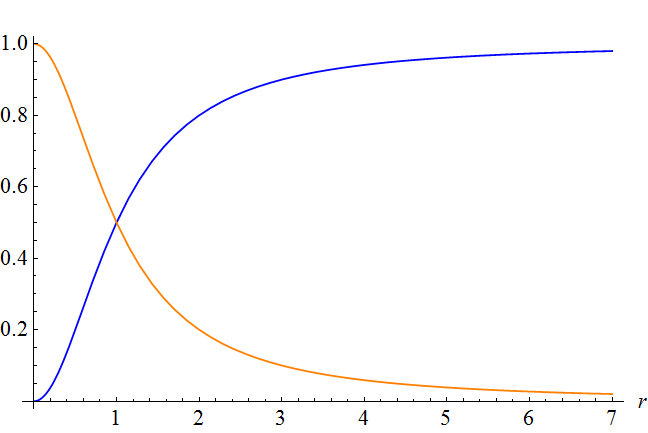}
\subcaption{Plot of $\Phi_+/M_0$ (blue) and $\Phi_-/M_0$ (yellow) as
  functions of $r\in [0,+\infty)$. The value $r=0$ corresponds to the
    center of $\mD$, while $r=+\infty$ corresponds to the plane end.
    Only the potential $\Phi_+$ leads to $\alpha$-attractor behavior
    when inflation takes place near the plane end.}
\label{fig:PhipmDiskr}
\end{minipage}
\caption{The potentials $\Phi_\pm$ on the Poincar\'e
  disk.}
\end{figure}

In polar coordinates $(\rho,\theta)$, the extended potentials
\eqref{hPhipm} of Section \ref{sec:Laplace} correspond to the
following globally well-behaved scalar potentials on $\mD$:
\ben
\label{Phipm}
\Phi_+=M_0\frac{r^2}{1+r^2}=M_0\frac{\left(\log\frac{1+\rho}{1-\rho}\right)^2}{1+\left(\log\frac{1+\rho}{1-\rho}\right)^2}~~,~~\Phi_-=\frac{M_0}{1+r^2}=\frac{M_0}{1+\left(\log\frac{1+\rho}{1-\rho}\right)^2}~~,
\een
where $\rho=|u|$. These potentials are shown in Figures
\ref{fig:PhipmDisk} and \ref{fig:PhipmDiskr}, which illustrate the
characteristic stretching toward the end when the potential is
expressed in semi-geodesic coordinates $(r,\theta)$ (with respect to
which the two locally-defined real scalar fields of the sigma model
have canonical kinetic terms). The supremum of $\Phi_+$ corresponds to
$\rho\rightarrow 1$ (being equal to $M_0$), while the infimum is
attained at $\rho=0$ (where $\Phi_+$ vanishes). On the other hand,
$\Phi_-$ tends to its vanishing infimum for $\rho\rightarrow 1$ and
has a maximum at $\rho=0$ (where it equals $M_0$). Notice that only
$\Phi_+$ leads to standard $\alpha$-attractor behavior.

\noindent Using the relation
$\sin\psi=2\sin(\psi/2)\cos(\psi/2)=\frac{2r}{1+r^2}$, the choice
\eqref{hPhi0} gives the following potential on $\mD$:
\be
\Phi_0=M_0\left[1+\frac{2r}{1+r^2}\cos\theta\right]=M_0\left[1+\frac{2 \log\frac{1+\rho}{1-\rho}}{1+\left(\log\frac{1+\rho}{1-\rho}\right)^2}\cos\theta\right]~~.
\ee

\section{The hyperbolic punctured disk}
\label{subsec:puncdisk}

\noindent The hyperbolic punctured disk $\mD^\ast$ (also known as the ``parabolic
cylinder'' \cite{Borthwick}) is the simplest example of a hyperbolic
surface with a cusp end. It also has a horn end.

\subsection{The hyperbolic metric}

\noindent
The punctured unit disk $\mD^\ast=\{u\in \C\,|\,0<|u|<1\}$ admits a
unique complete hyperbolic metric given by \cite{BM}:
\ben
\label{gDast}
\dd s_{\mD^\ast}^2=\lambda_{\mD^\ast}^{2}(u,\bar{u})|\dd u|^2~~,
~\mathrm{where}~~\lambda_{\mD^\ast}(u,\bar{u})=\frac{1}{|u|\log(1/|u|)}~~.
\een
In particular, $\Re u$ and $\Im u$ are isothermal coordinates. In
polar coordinates $(\rho,\theta)$ defined through:
\be
u=\rho e^{\i \theta}~~(\rho=|u|\in (0,1))~~,
\ee
the metric takes the form: 
\be
\dd s_{\mD^\ast}^2=\frac{1}{(\rho\log\rho)^2}(\dd \rho^2+\rho^2 \dd \theta^2)~~.
\ee
The center of $\mD^\ast$ corresponds to the cusp end, while the bounding
circle of $\mD^\ast$ corresponds to the horn end (see below). The Euclidean
circle at $\rho=e^{-2\pi}$ is a horocycle of hyperbolic length $1$. Notice that
$\mD^\ast$ has infinite hyperbolic area.

\subsection{Diffeomorphism to the punctured plane}

\noindent
One can introduce an orthogonal coordinate system $(\fr,\theta)\in
(0,+\infty)\times (0,2\pi)$ on $\mD^\ast$ through the coordinate
transformation:
\ben
\label{rhomDast}
\fr \eqdef \frac{1}{\log\left(\frac{1}{\rho}\right)}=\frac{1}{|\log\rho|}\in(0,+\infty)~~\mathrm{i.e.}~~\rho=e^{-\frac{1}{\fr}}\in (0,1)~~.
\een
This gives a diffeomorphism between $\mD^\ast$ and the punctured
complex plane $\C^\ast\eqdef\C\setminus \{0\}$ with complex coordinate:
\be
\zeta=\fr e^{\i\theta}~~.
\ee
In this coordinate system, the metric \eqref{gDast} takes the form:
\ben
\label{cmetric1}
\dd s_{\mD^\ast}^2=\frac{1}{\fr^2} \dd\fr^2+\fr^2\dd\theta^2~~.
\een
The center of $\mD^\ast$ corresponds to $\fr\rightarrow 0$, while the
bounding circle of $\mD^\ast$ corresponds to $\fr\rightarrow +\infty$.

\subsection{Partial isometric embedding into Euclidean 3-space}

\begin{figure}[H]
\centering \includegraphics[width=50mm]{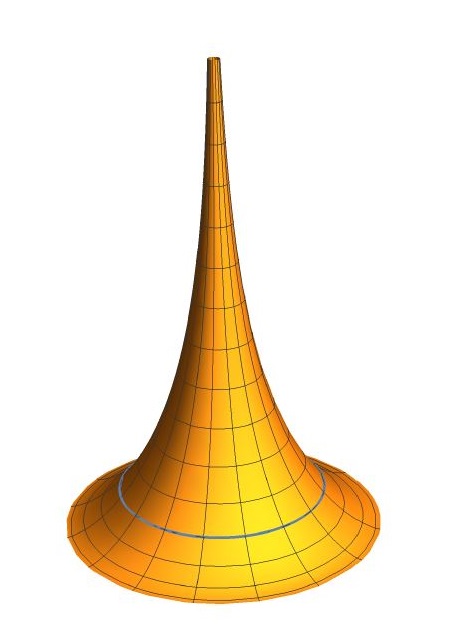}
\caption{Upper half of a tractricoid (pseudosphere). The horocycle of
  length one is drawn in light blue. It separates the half-tractricoid
  into a cusp (upper part) and a portion of a horn (lower part). The
  latter is bounded by a horocycle of length $2\pi$.}
\label{Tractroid}
\end{figure}

\noindent
One can isometrically embed the portion $0<|u|<\frac{1}{e}$ of the
hyperbolic punctured disk into Euclidean $\R^3$ as the open
half-tractricoid\footnote{The surface obtained by revolving an open
  half of a tractrix along its asymptote \cite{Kuhnel}.} $\cE$ defined
in cylindrical coordinates $(\fr,\theta,x_3)$ (where
$x_1=\fr\cos\theta$ and $x_2=\fr\sin\theta$) by the parametric
equations:
\be
\fr=\frac{1}{\cosh(\ft)}\in (0,1)~~,~~x_3=\ft-\tanh(\ft)~~(\ft\in (0,+\infty))~~.
\ee
Indeed, it is easy to see that the Euclidean metric of $\R^3$ induces
a metric on $\cE$ which coincides with \eqref{cmetric1}. This is the
classical pseudosphere model of Beltrami (see Figure \ref{Tractroid}).

\subsection{The end compactification of $\mD^\ast$}

\noindent
The stereographic projection \eqref{sterproj} identifies $\rS^2$ with
the one-point compactification of the $\zeta$-plane. This shows
explicitly that $\rS^2$ is the end compactification of $\mD^\ast$,
where the north pole $\psi=0$ corresponds to the horn end and the
south pole $\psi=\pi$ corresponds to the cusp end. The embedding
$j:\mD^\ast\rightarrow \rS^2$ is given by:
\be
\psi=2\arctan(|\log\rho|)~~,~~\theta=\theta~~.
\ee

\subsection{Semi-geodesic coordinates}

\noindent
The further change of variables:
\ben
\label{fr}
\fr=\frac{1}{2\pi}e^{-r}~~\mathrm{i.e.}~~r=-\log(2\pi \fr)=\log\left(\frac{|\log \rho|}{2\pi}\right)\in (-\infty,+\infty)
\een
brings the metric \eqref{cmetric1} to the form:
\ben
\label{cmetric2}
\dd s_{\mD^\ast}^2=\dd r^2+\frac{e^{-2r}}{(2\pi)^2}\dd\theta^2~~,
\een
where $r\in \R$. In particular, $(r,\theta)$ are semi-geodesic
coordinates. The center $u=0$ of $\mD^\ast$ corresponds to
$r\rightarrow +\infty$ while the bounding circle $|u|=1$ of $\mD^\ast$
corresponds to $r\rightarrow -\infty$. The horocycle at $r=0$
(i.e. $\rho=e^{-2\pi}$) has length $1$.

\subsection{The hyperbolic cusp}

\noindent Let: 
\ben
\label{kappa}
\kappa \eqdef e^{-2\pi}~~.
\een
As mentioned above, the Euclidean circle $|u|=\kappa$ has hyperbolic
length $1$. The {\em hyperbolic cusp} (cf. \cite{genalpha})
corresponds to the portion of $\mD^\ast$ lying inside this circle (see
Figure \ref{Tractroid}), which is the open punctured disk:
\ben
\label{mDastr0}
\rC\eqdef \{u\in \C\,\, |\,\, 0<|u|<\kappa\}\subset \mD^\ast~~,
\een
endowed with the restriction of the metric \eqref{gDast}; notice that
the restricted metric is not complete. In coordinates $(\fr,\theta)$, the
metric on $\rC$ is obtained by restricting \eqref{cmetric1} to the
range $\fr\in (0,\fr_0)$, where:
\be
\fr_0\eqdef \frac{1}{2\pi}~~.
\ee
In semi-geodesic coordinates $(r,\theta)$ the cusp metric is given by
\eqref{cmetric2} with the restriction $r\in (0,+\infty)$. Notice 
that $\rC$ has hyperbolic area equal to $1$.

\subsection{The hyperbolic horn}

\noindent
By definition, the {\em hyperbolic horn} is the annulus:
\be
\rH=\{u\in \C\,\, |\,\, \kappa <|u|<1\}\subset \mD^\ast~~,
\ee
endowed with the (incomplete) restriction of the metric
\eqref{gDast}. In coordinates $(\fr,\theta)$, the horn metric is
obtained from \eqref{cmetric1} by restricting the range of $\fr$ to
$(\fr_0,+\infty)$. In coordinates $(r,\theta)$, the metric takes
the form \eqref{cmetric2}, with the restriction $r\in
(-\infty,0)$. Defining $r'\eqdef -r$, this can be brought to the form:
\ben
\label{hmetric2}
\dd s^2_{\rH}=(\dd r')^2+\frac{e^{2r'}}{(2\pi)^2}\dd\theta^2~~,~~\mathrm{with}~~r'\in (0,+\infty)~~,
\een
where the bounding circle $|u|=1$ of $\mD^\ast$ corresponds to
$r'\rightarrow +\infty$.

\subsection{Canonical uniformization to $\H$}

\noindent
The punctured disk is uniformized to the Poincar\'e half-plane
\eqref{PoincareHP} with complex coordinate $\tau$ by the parabolic
cyclic group $\Gamma_P\eqdef\langle P\rangle\subset \PSL(2,\R)$
generated by the translation:
\ben
\label{trsl}
\tau \rightarrow \tau+1~~,
\een
which corresponds to the parabolic element: 
\ben
\label{P0}
P \eqdef \left[\begin{array}{cc} 1 & 1\\ 0 & 1\end{array}\right]\in \PSL(2,\R)~~.
\een
This fixes the point $\tau=\infty\in \partial_\infty \H$. The
uniformization map is:
\ben
\label{pipuncdiskcan}
u=\pi_\H(\tau)=e^{2\pi \i \tau}~~.
\een
The hyperbolic cusp $\rC$ is the projection through $\pi_\H$ of the
cusp domain:
\be
\cC_\H=\{\tau\in \H|\Im \tau >1\}~~,
\ee
which is bounded by the horocycle: 
\be
c_\H=\{\tau\in \H|\Im \tau=1\}~~.
\ee
This horocycle is tangent to the conformal boundary of $\H$ at the
point $\tau=\infty$, which projects through $\pi_\H$ to the cusp ideal
point of the end compactification of $\mD^\ast$.

A fundamental polygon for the action of $\Gamma_P$ on $\H$ is
given by the semi-infinite vertical strip:
\ben
\label{Dpuncdisk}
\fD_\H=\left\{\tau \in \H \, |\, 0 <  \Re \tau < 1\right\}~~,
\een
and has vertices at the points (see Figure \ref{fig:mDast1}):
\ben
\label{ABCPuncDisk}
A:\{\tau=\infty\}~,~B:\{\tau=0\}~,~C:\{\tau=1\}~~.
\een
The Poincar\'e side pairing maps $(AB)$ into $(AC)$ through the
transformation \eqref{trsl}, which generates $\Gamma_P$. The relative
cusp neighborhood \cite{genalpha} with respect to $\fD_\H$ is the
intersection $\fC_\H\eqdef \cC_\H\cap \fD_\H$. A lifted scalar potential
$\tPhi=\Phi\circ \pi_\H$ is given by:
\be
\tPhi(\tau)=\Phi(e^{2\pi i \tau})~~,
\ee
being invariant under the translation \eqref{trsl}. In particular, the
restrictions of $\tPhi$ to the sides $(AB)$ and $(AC)$ agree through
the Poincar\'e pairing.

\subsection{Canonical uniformization to $\mD$}

\begin{figure}[H]
\centering
\begin{minipage}{.47\textwidth}
\centering
\includegraphics[width=0.8\linewidth]{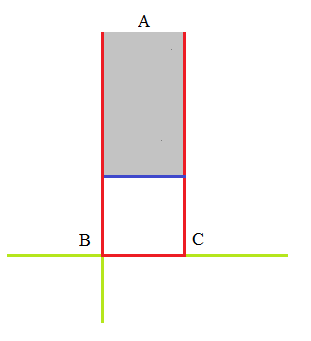}
\subcaption{A fundamental polygon $\fD_\H$ for the punctured disk on
  $\H$. The relative cusp neighborhood $\fC_\H$ of the vertex $A$
  corresponds to the shaded region.}
\label{fig:mDast1}
\end{minipage}\hfill
\begin{minipage}{.47\textwidth}
\centering
\vspace{-3mm}
\includegraphics[width=0.87\linewidth]{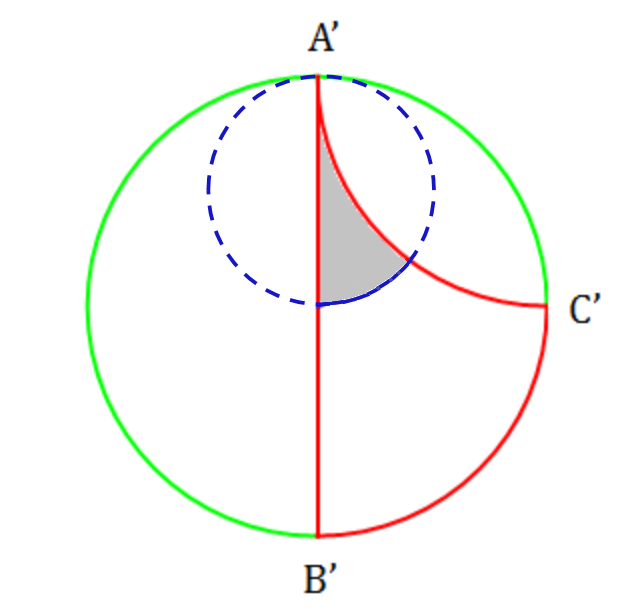}
\subcaption{A fundamental polygon $\fD_\mD$ for the punctured disk on
  $\mD$. The shaded region is the relative cusp neighborhood $\fC_\mD$
  of the vertex $A'$.}
\label{fig:mDast2}
\end{minipage}
\caption{Fundamental polygons for the uniformization of $\mD^\ast$ to
  the Poincar\'e half-plane and to the hyperbolic disk.}
\end{figure}

\noindent
For completeness and comparison with \cite{genalpha}, we also give
the canonical uniformization of $\mD^\ast$ to the hyperbolic disk.
The Cayley transformation:
\ben
\label{Cayley}
u=\frac{\i-\tau}{\i \tau-1}
\een
is an isometry from $\H$ to $\mD$. When uniformizing $\mD^\ast$ to
$\mD$, the fundamental polygon $\fD_\H$ becomes a hyperbolic triangle
$\fD_{\mD}$ with vertices at the following points, which
correspond respectively to the points $A$, $B$ and $C$ of \eqref{ABCPuncDisk} 
through the Cayley transformation:
\be
A':\{u=\i\}~,~B':\{u=-\i\}~,~~C':\{u=1\} 
\ee
and a free side connecting the points $B'$ and $C'$ (see Figure
\ref{fig:mDast2}). The sides of this triangle are a segment connecting
$A'$ to $B'$ (which passes through the origin of $\mD$), the portion of
$\partial_\infty\mD$ connecting $B'$ to $C'$ and a Euclidean circular
arc orthogonal to $\partial_\infty \mD$ which connects $A'$ to
$C'$. The hyperbolic cusp neighborhood $\cC_\mD$ of the vertex $A'$ is bounded
by the horocycle:
\be
c_{\mD}=\left\{ \frac{t}{2-\i t}\in \fD_{\mD} \, |\, t\in \R \right\}~~,
\ee
which is tangent to $\partial_\infty \mD$ at the point $A'$. The
intersection of $\cC_\mD$ with $\fD_\mD$ is the relative cusp
neighborhood $\fC_\mD$ with respect to $\fD_\mD$ (see \cite{genalpha}), 
which is the image of $\fC_\H$ through the Cayley transformation. 

\subsection{Globally well-behaved scalar potentials on $\mD^\ast$}

\begin{figure}[H]
\centering
\vskip -0.5em
\begin{minipage}{.47\textwidth}
\vskip 0.5em
\centering \includegraphics[width=1\linewidth]{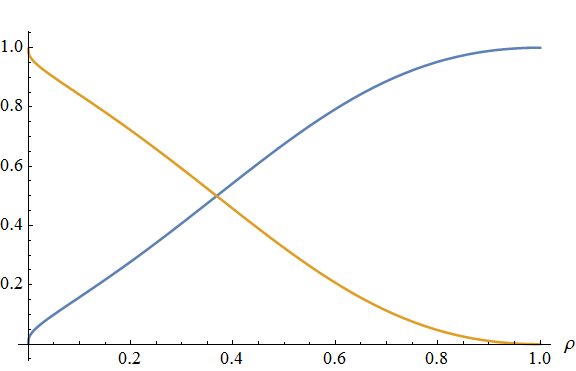}
\vskip 1em
\subcaption{Plot of $\Phi_+/M_0$ (blue) and $\Phi_-/M_0$ (yellow) as
  functions of $\rho\in (0,1)$. The value $\rho=0$ corresponds to the
  cusp end, while $\rho=1$ corresponds to the horn end.}
\label{fig:PhipmPuncturedDisk}
\end{minipage}\hfill
\begin{minipage}{.47\textwidth}
\centering \includegraphics[width=0.9\linewidth]{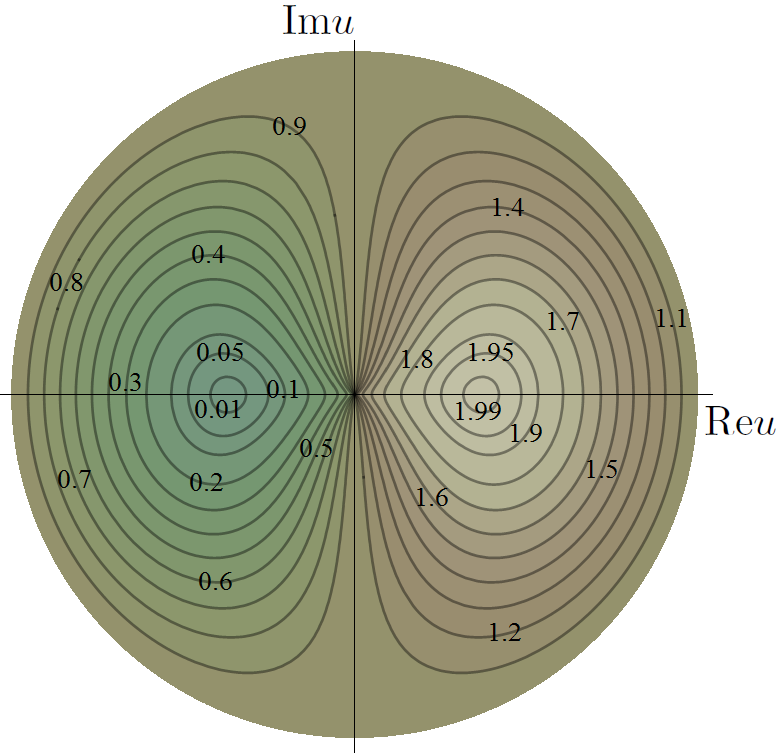}
\subcaption{Level plot of $\Phi_0/M_0$ on the punctured disk. Darker
  tones indicate lower values of $\Phi_0/M_0$.}
\label{fig:LevelPlotPuncturedDisk}
\end{minipage}
\vskip 1em
\caption{The potentials $\Phi_\pm$ and $\Phi_0$ on the hyperbolic punctured disk.}
\label{fig:PhipmPuncturedDiskAll}
\end{figure}

\noindent
A scalar potential $\Phi:\Sigma\rightarrow \R$ is globally
well-behaved on $\mD^\ast$ iff there exists a smooth map
$\hPhi:\rS^2\rightarrow \R$ such that:
\be
\Phi(\fr,\theta)=\hPhi(2\arccot(\fr),\theta)~~,
\ee
i.e.: 
\be
\Phi(\rho,\theta)=\hPhi(2\arctan(|\log\rho|),\theta)~~,
\ee
where $\rho=|u|$. Expansion \eqref{hPhiExp} gives the
uniformly-convergent series:
\be
\Phi(\rho,\theta)=\sum_{l=0}^\infty \sum_{m=-l}^l D_{lm} Y_{lm}(2\arctan(|\log\rho|),\theta)~~.
\ee
For the choices \eqref{hPhipm} and \eqref{hPhi0}, we find: 
\beqan
\label{PuncDiskPot}
&& \Phi_+=M_0\frac{\fr^2}{1+\fr^2}=M_0\frac{1}{1+(\log\rho)^2}\nn\\
&& \Phi_-=M_0\frac{1}{1+\fr^2}=M_0\frac{(\log\rho)^2}{1+(\log\rho)^2}\\
&& \Phi_0=M_0\left[1+\frac{2\fr\cos\theta}{1+\fr^2}\right]=M_0\left[1+ \frac{2|\log \rho|}{1+(\log\rho)^2}\cos\theta\right]~~.\nn
\eeqan
These potentials are shown in Figure
\ref{fig:PhipmPuncturedDiskAll}. Notice that $\Phi_+$ leads to
$\alpha$-attractor behavior if inflation takes place near the horn
end, while $\Phi_-$ leads to $\alpha$-attractor behavior if inflation
takes place near the cusp end \cite{genalpha}. The extended potentials
$\hPhi_\pm$ have maxima and minima at the two ideal points of the end
compactification of $\mD^\ast$ (which correspond to the north and
south poles of $\rS^2$). On the other hand, $\hPhi_0$ does {\em not}
have extrema at the ideal points; its extrema coincide with those of
$\Phi_0$, being located inside $\mD^\ast$. The minimum (equal to zero)
is at the point $u=-\frac{1}{e}\approx -0.36$ while the maximum (equal
to $2M_0$) is at $u=+\frac{1}{e}\approx +0.36$.

\paragraph{Remark.} 
Using \eqref{fr}, we find the following expressions in semi-geodesic coordinates: 
\ben
\label{PhipmDast}
\Phi_+(r)=\frac{M_0}{1+(2\pi)^2 e^{2r}}~~,~~\Phi_-(r)=\frac{M_0}{1+\frac{e^{-2r}}{(2\pi)^2}}~~,~~\Phi_0(r,\theta)=M_0\left[1+\frac{4\pi e^{-r}\cos\theta}{(2\pi)^2+e^{-2r}}\right]~~.
\een

\paragraph{Lift of the potentials $\Phi_\pm$ and $\Phi_0$ to $\H$.}

\begin{figure}[H]
\centering
\begin{minipage}{.47\textwidth}
\centering
\vskip 5mm
\includegraphics[width=1\linewidth]{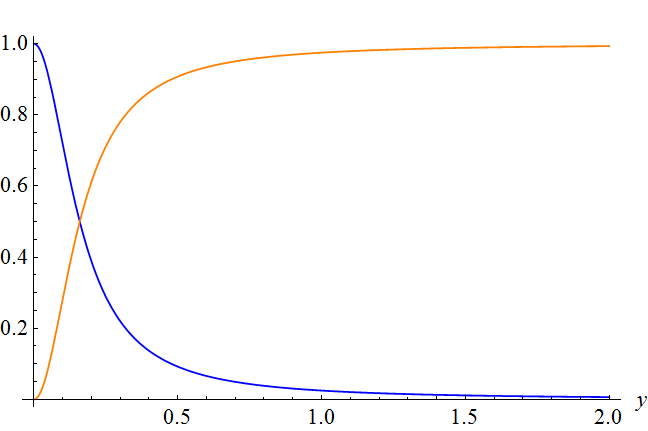}
\vskip 3em \subcaption{Plot of $\tPhi_+/M_0$ (blue) and $\tPhi_-/M_0$
  (yellow) as functions of $y=\Im \tau\in (0,+\infty)$. The values
  $y=0$ and $y=+\infty$ correspond respectively to the horn and cusp
  ends.}
\label{fig:PhipmPuncturedDisky}
\end{minipage}
\hfill
\begin{minipage}{.47\textwidth}
\centering \includegraphics[width=1\linewidth]{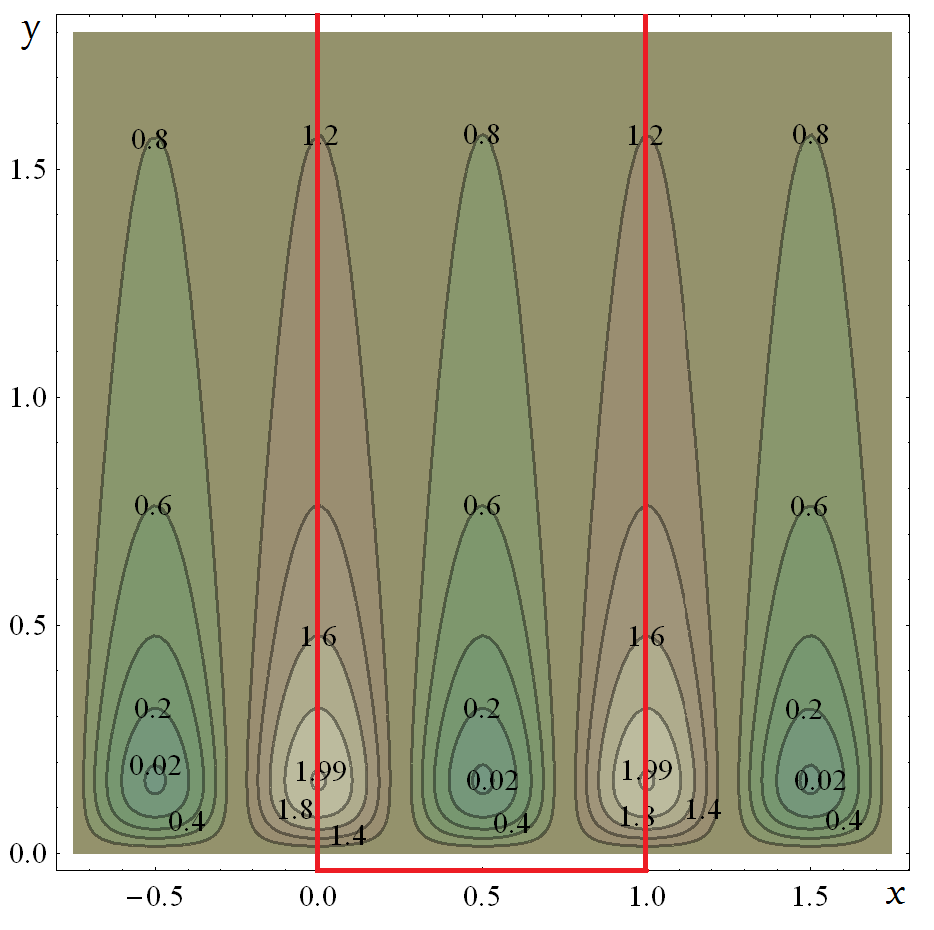}
\subcaption{Level plot of $\tPhi_0/M_0$. Higher values of $\tPhi_0$
  correspond to lighter tones. The fundamental domain $\fD_\H$ is
  bounded by the red lines.}
\label{fig:LevelSets}
\end{minipage}
\caption{The lifted potentials $\tPhi_\pm$ and $\tPhi_0$.}
\label{fig:PuncDiskLiftedPotAll}
\end{figure}

Consider the well-behaved scalar potentials $\Phi_\pm$ and $\Phi_0$ on
$\mD^{\ast}$ given in \eqref{PuncDiskPot}. Let $x=\Re\tau$ and
$y\eqdef \Im \tau$. Then the covering map \eqref{pipuncdiskcan} reads:
\be
u=\pi_{\H}(\tau) = e^{2\pi \i\tau}=e^{-2\pi y}[\cos (2\pi x)+\i \sin (2\pi x)]~~,
\ee
which gives:
\ben
\label{PuncDiskRhoTheta}
\rho=|u|=e^{-2\pi y}~~,~~\theta=\arg(u)=2\pi x~(\mathrm{mod}\, 2\pi)~~. 
\een
Hence the potentials $\Phi_\pm$ and $\Phi_0$ have the following lifts
to $\H$ (see Figure \ref{fig:PuncDiskLiftedPotAll}):
\ben
\label{tPhipmDasty}
\tPhi_+= M_0\frac{1}{1+(2\pi)^2 y^2}~~,~~\tPhi_- = M_0\frac{y^2}{(2\pi)^{-2} + y^2}~~,~~\tPhi_0=M_0 \left[1+\frac{4\pi y \cos(2\pi x)}{1+4\pi^2 y^2}\right]~~.
\een
The function $\tPhi_0(\tau)$ (which is periodic under $\tau\rightarrow
\tau+1$) attains its minimum $\min\tPhi_0=0$ at the points
$\tau=n+\frac{1}{2}+\frac{\i}{2\pi}\approx n+0.5+0.16 \i$ (with $n\in
\Z$), while the maximum $\max\tPhi_0=2M_0$ is attained for
$\tau=n+\frac{\i}{2\pi}\approx n+0.16 \i$ with $n\in \Z$.

\subsection{Cosmological trajectories on the hyperbolic punctured disk}
\label{subsec:TrajPuncDisk}

\noindent In this subsection, we present examples of
numerically-computed trajectories on $\mD^\ast$ for the vanishing
scalar potential and for the globally well-behaved scalar potentials
$\Phi_\pm$ and $\Phi_0$. These were obtained as explained in
Subsection \ref{subsec:lift}, by numerically computing solutions of
the system \eqref{elplane0} on the Poincar\'e half-plane for the
corresponding lifted potentials and then projecting these trajectories
to the hyperbolic punctured disk using the explicitly-known
uniformization map \eqref{pipuncdiskcan} (which is equivalent with
\eqref{PuncDiskRhoTheta}).

\paragraph{Trajectories for vanishing scalar potential.}

To understand the effect of the hyperbolic metric on the dynamics, we
start with the case of a vanishing scalar potential $\Phi=0$. Then
$\tPhi=0$ and one immediately checks that straight lines given by
constant functions $x(t)=x_0, y(t)=y_0$ are solutions of
\eqref{elplane} for any initial point $(x_0,y_0)\in \H$, with initial
velocity zero. This means that a scalar field starting ``at rest''
remains at rest for all times. On the other hand, numerical
computation shows that any solution of \eqref{elplane} tends to the
real axis for $t\rightarrow +\infty$, irrespective of its initial
conditions.  As a consequence, any solution defined on $\mD^\ast$
(which is obtained by projecting a solution defined on $\H$ through
the map \eqref{pipuncdiskcan}) will tend toward the horn end as
$t\rightarrow +\infty$.  This shows that the hyperbolic metric acts as
an effective force which repulses $\varphi$ away from the cusp
end. Notice that a global trajectory defined on $\mD^\ast$ can have
cusp and self-intersection points and hence that it need not
correspond to an embedded curve in $\mD^\ast$.

\begin{figure}[H]
\centering
\begin{minipage}{.5\textwidth}
\centering \includegraphics[width=.85\linewidth]{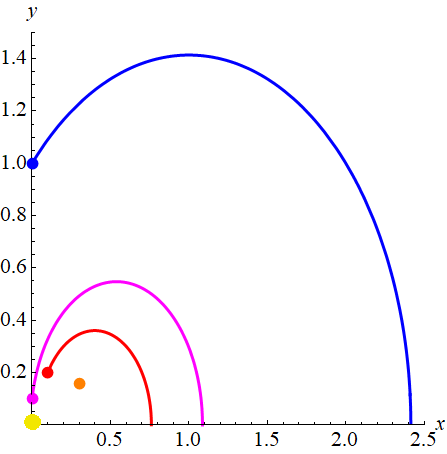}
\vskip 1em \subcaption{Trajectories for $\tPhi=0$ on the Poincar\'e
  half-plane. The solutions shown in orange and yellow are stationary.}
\label{fig:PuncDiskNoPot}
\end{minipage}\hfill
\begin{minipage}{.47\textwidth}
\centering \includegraphics[width=.95 \linewidth]{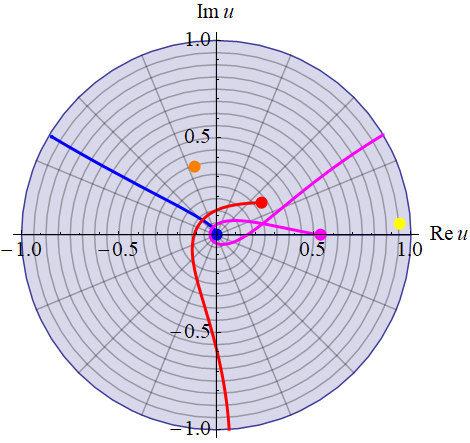}
\subcaption{Projection of the trajectories shown at the left 
to the hyperbolic punctured disk.}
\label{fig:PuncDiskNoPotProj}
\end{minipage}
\caption{Numerical solutions for $\Phi=0$ and $\alpha=\frac{M_0}{3}$.}
\label{fig:PuncDiskNoPotAll}
\end{figure}

Figure \ref{fig:PuncDiskNoPotAll} shows five trajectories on $\H$ and their 
projections to $\mD^\ast$
for $\Phi=0$ and $\alpha=\frac{M_0}{3}$, with the following initial
conditions:

\begin{table}[H]
\centering
\begin{tabular}{|c|c|c|}
\toprule
trajectory & $\tau_0$ & $\tv_0$ \\
\midrule\midrule
 orange  & $0.3+0.159\,\i$  & $0$ \\
\hline
yellow & $0.01+0.009\,\i$  & $0$\\
\hline
 red  & $0.1+0.2\,\i$ & $2+3\,\i$  \\
\hline
blue & $\i$ & $1+\,\i$  \\
\hline
 magenta & $0.1\,\i$  & $1.3+7\,\i$\\

\bottomrule
\end{tabular}
\caption{Initial conditions for five trajectories on the Poincar\'e half-plane.}
\label{table:InCond}
\end{table}

\paragraph{Trajectories  for $\Phi_-$.}

\begin{figure}[H]
\centering
\begin{minipage}{.47\textwidth}
\centering
\vskip -1em
\includegraphics[width=.9\linewidth]{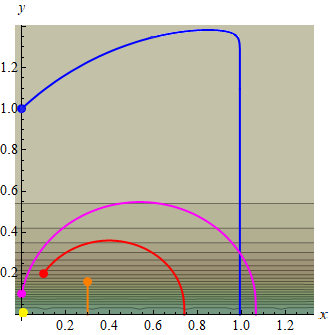}
\subcaption{Trajectories for $\tPhi=\tPhi_-$ on the Poincar\'e half-plane, drawn  
over a  level plot of $\tPhi_-$ on $\H$.}
\label{fig:PuncDiskPhiMinus}
\end{minipage}\hfill
\begin{minipage}{.47\textwidth}
\centering
\includegraphics[width=.9\linewidth]{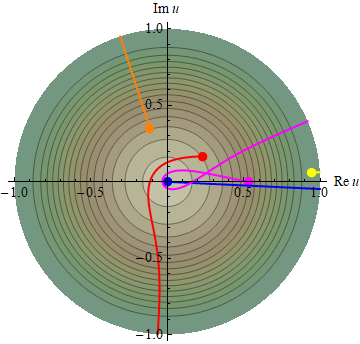}
\vskip 2mm
\subcaption{Projection of the trajectories shown at the left to the hyperbolic punctured disk.
We also show a level plot of $\Phi_-$ on $\D^\ast$.}

\label{fig:PuncDiskPhiMinusProj}
\end{minipage}

\caption{Numerical solutions for $\Phi=\Phi_-$ and $\alpha=\frac{M_0}{3}$.}
\label{fig:PuncDiskPhiMinusAll}
\end{figure}

\noindent Five  lifted trajectories (and their projections to $\mD^\ast$)
for $\alpha=\frac{M_0}{3}$ and $\Phi=\Phi_-$ with the initial
conditions given in Table \ref{table:InCond} are shown in Figure
\ref{fig:PuncDiskPhiMinusAll}. Since $\hPhi_-$ has a maximum at the
cusp end (center of the disk) and a minimum at the horn end, it
reinforces the effect of the hyperbolic metric, together with which it
produces an effective repulsion away from the cusp end. In particular,
the two trajectories which start with vanishing initial velocity are no longer
stationary, but evolve for $t\rightarrow +\infty$ to the funnel end
(see the orange and yellow trajectories in the figures). Any other trajectory
(irrespective of its initial velocity) also evolves for $t\rightarrow
+\infty$ to the funnel end.

\paragraph{Trajectories  for $\Phi_+$.}

\begin{figure}[H]
\centering
\begin{minipage}{.47\textwidth}
\centering
\includegraphics[width=.9\linewidth]{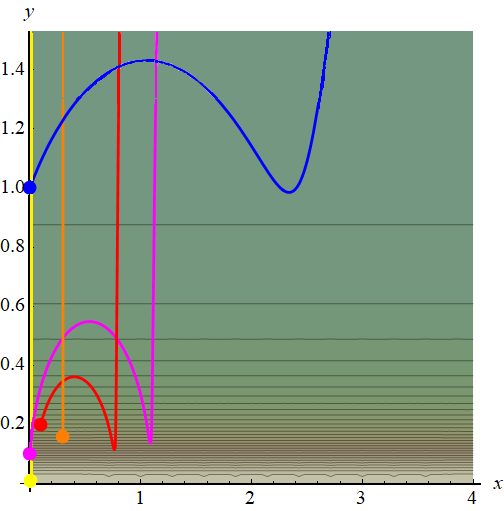}
\subcaption{Trajectories for $\tPhi=\tPhi_+$ on the Poincar\'e half-plane
and a level plot of $\Phi_+$ on $\H$.}
\label{fig:PuncDiskPhiPlus}
\end{minipage}\hfill
\begin{minipage}{.47\textwidth}
\centering
\includegraphics[width=.9\linewidth]{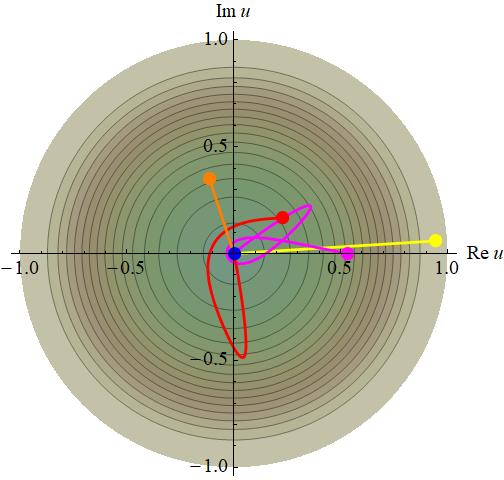}
\vskip 2mm
\subcaption{Projection of the trajectories shown at the left to the hyperbolic punctured disk
and a level plot of $\tPhi_+$ on $\D^\ast$.}
\label{fig:PuncDiskPhiPlusProj}
\end{minipage}
\caption{Numerical solutions for $\Phi=\Phi_+$ and $\alpha=\frac{M_0}{3}$.}
\label{fig:PuncDiskPhiPlusAll}
\end{figure}

\noindent Five lifted trajectories for $\alpha=\frac{M_0}{3}$ and
$\Phi=\Phi_+$ (and their projections to $\mD^\ast$) with the initial
conditions given in Table \ref{table:InCond} are shown in Figure
\ref{fig:PuncDiskPhiPlusAll}. Since $\hPhi_+$ has a minimum at the
cusp end (which corresponds to the center of the disk), it produces an
attractive force towards the cusp end, which acts as a counterbalance
to the repulsive effect of the hyperbolic metric.

\paragraph{Trajectories for $\Phi_0$.}

\begin{figure}[H]
\centering
\begin{minipage}{.47\textwidth}
\centering \includegraphics[width=1\linewidth]{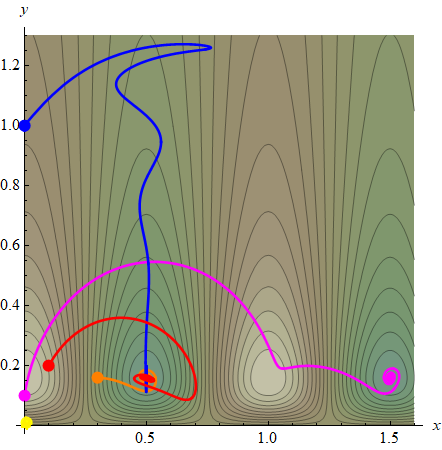}
\vskip 1mm\subcaption{Trajectories for $\tPhi=\tPhi_0$ on the Poincar\'e
  half-plane.}
\label{fig:PuncDiskPhi0}
\end{minipage}\hfill
\begin{minipage}{.47\textwidth}
\vskip 1mm
\centering \includegraphics[width=1.05\linewidth]{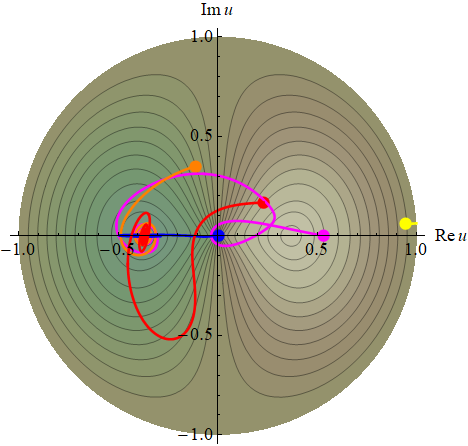}
\vskip 2mm \subcaption{Projection of the trajectories shown at the
  left to the hyperbolic punctured disk.}

\label{fig:PuncDiskPhi0Proj}
\end{minipage}
\caption{Numerical solutions for $\Phi=\Phi_0$ and
  $\alpha=\frac{M_0}{3}$.}
\label{fig:PuncDiskPhi0All}
\end{figure}

Five lifted trajectories (and their projections to $\mD^\ast$) 
for $\alpha=\frac{M_0}{3}$ and $\Phi=\Phi_0$
with the initial conditions given in Table \ref{table:InCond} are
shown in Figure \ref{fig:PuncDiskPhi0All}.

\begin{figure}[H]
\centering
\begin{minipage}{.47\textwidth}
\centering \includegraphics[width=.88\linewidth]{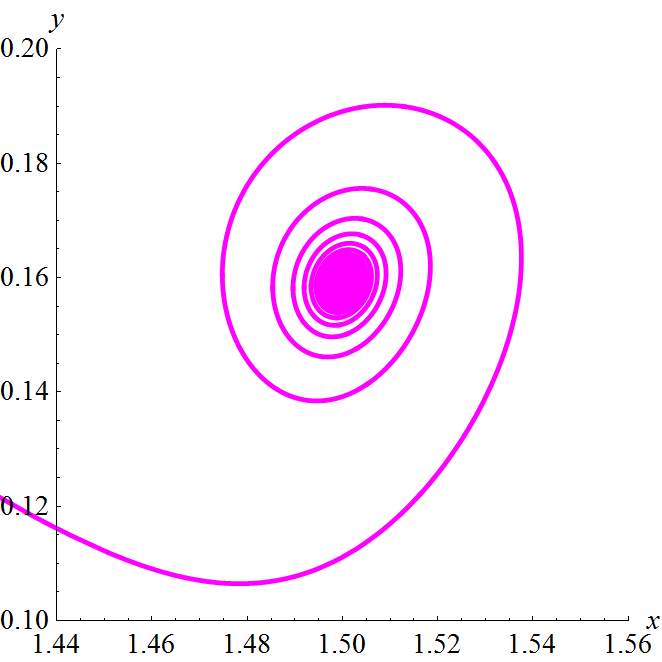}
\subcaption{Detail of the magenta trajectory shown in Figure
  \ref{fig:PuncDiskPhi0} near one minimum of $\tPhi_0$.}
\label{fig:TrajDetail}
\end{minipage}\hfill
\begin{minipage}{.47\textwidth}
\centering \includegraphics[width=.9\linewidth]{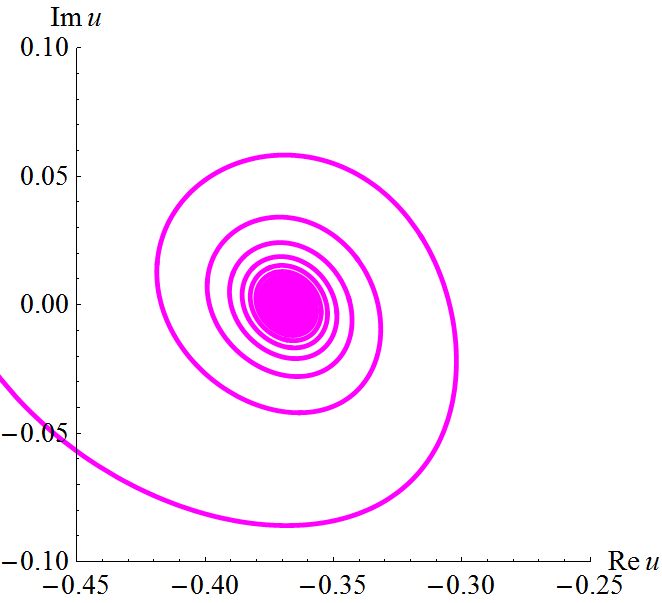}
\vskip 1.0mm \subcaption{Projection of the trajectory shown at the
  left to the hyperbolic punctured disk.}
\vskip 3.0mm
\label{fig:TrajDetailProj}
\end{minipage}
\caption{Detail of the magenta trajectory.}
\label{fig:TrajDetailAll}
\end{figure}

\noindent Figure \ref{fig:TrajDetailAll} shows in more detail the
behavior of the magenta trajectory of Figure \ref{fig:PuncDiskPhi0}
near the local minimum of $\tPhi_0$ located at
$\tau=\frac{3}{2}+\frac{\i}{2\pi}$ (which projects to
$u=-\frac{1}{e}$) and of its projection to $\mD^\ast$. For this
solution, $\varphi$ evolves in a spiral around the minimum of
$\Phi_0$, until it settles at the minimum.

\subsection{Inflationary regions and number of e-folds}
\label{subsec:edisk}

\noindent Recall from \cite{modular} the expressions for 
the Hubble parameter and the critical Hubble parameter:
\ben
\label{Hubble}
H(t)=\frac{1}{3M_0}\sqrt{3\alpha \frac{\dot{x}(t)^2+\dot{y}(t)^2}{y(t)^2}+2\tPhi(x(t),y(t))}~~,
\een
\ben
\label{critHubble}
H_c(t)=\frac{1}{M_0}\sqrt{\frac{\tPhi(t)}{3}}~.
\een
For the inflationary regions the following inequality should be satisfied:
\ben
\label{cond}
H(t)<H_c(t)~,
\een
while the number of e-folds ${ N}$ is given by integrating $H(t)$ over 
the first inflationary time interval $t_I$:
\ben
\label{efolds}
{ N}=\int_{0}^{t_I}H(t)\dd t~.
\een
Analyzing the five trajectories with the initial conditions given in
Table \ref{table:InCond}, we find that only the orange and yellow
trajectories start in inflationary regime for each of the three
potentials $\Phi_+,\Phi_-,\Phi_0$, while the other three trajectories
do not start in inflationary regime for any of these
potentials. We also notice that the yellow trajectory in  $\Phi_0$
  is in inflationary regime for all times.
Calculating the number of e-folds, we find for the orange
trajectory the values $0.44$ for the potential $\Phi_0$ and $0.75$ for
$\Phi_\pm$, while the yellow trajectory has $55$ e-folds for the
potential $\Phi_+$ (so it satisfies the observational requirement of
$50-60$ e-folds) and $0.3$ for $\Phi_-$. Varying the
initial conditions of the yellow trajectory in the potential $\Phi_+$
(namely varying $\Im\tau_0$ in the range $[0.0094,0.0096]$) produces
many other trajectories with $N$ lying in the observationally
expected range of $[50,60]$ e-folds.



\begin{figure}[H] \centering \includegraphics[width=60mm]{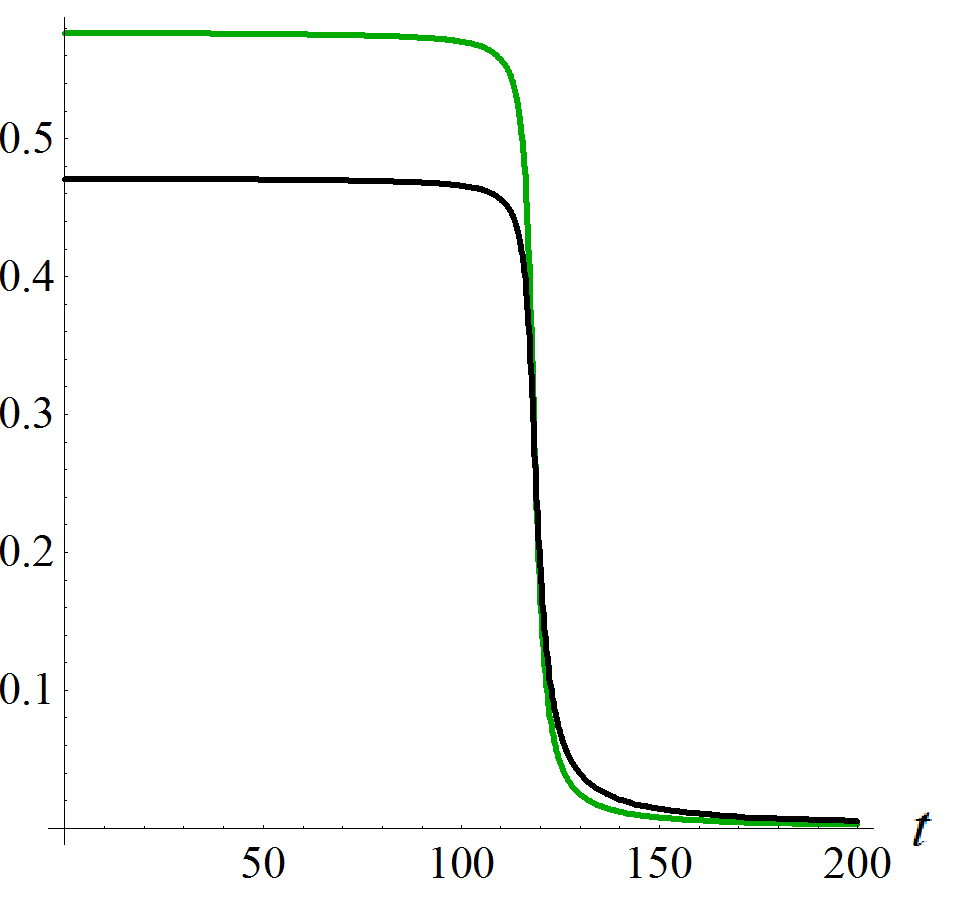}
\caption{Plot of $H(t)/\sqrt{M_0}$ (black) and $H_c(t)/\sqrt{M_0}$
(green) for the yellow trajectory in the potential $\Phi_+$. For
better clarity, we truncated the plot at $t=200 \,s$.}
\label{fig:NHubPDphiP}
\end{figure}

\begin{figure}[H]
\centering \includegraphics[width=50mm]{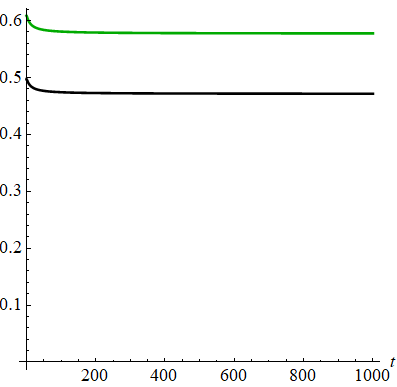}~\includegraphics[width=50mm]{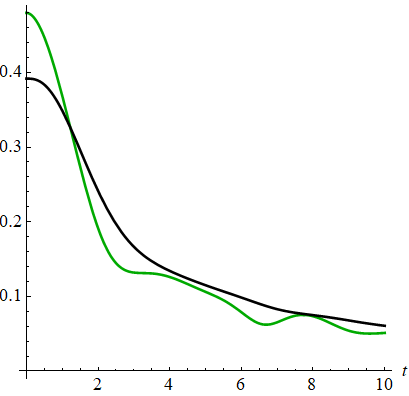}~\includegraphics[width=50mm]{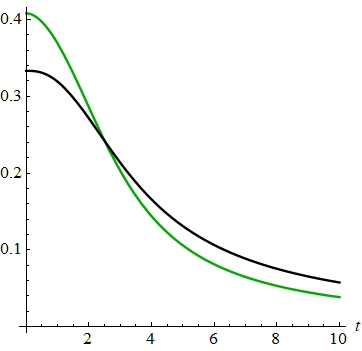}
\caption{Plots of $H(t)/\sqrt{M_0}$ (black) and $H_c(t)/\sqrt{M_0}$
(green) for the yellow trajectory in the potential $\Phi_0$ and for
the orange trajectory in the potentials $\Phi_0$ and $\Phi_-$,
respectively.  For better clarity, we truncated the plots at
convenient values of $t$.}
\label{fig:HubPDothers}
\end{figure}

\section{Hyperbolic annuli}
\label{subsec:annulus}

\noindent Hyperbolic annuli (also known as ``hyperbolic cylinders''
\cite{Borthwick}) have a single modulus and two funnel ends.

\subsection{The hyperbolic metric}

\noindent
Let $R>1$ be a real number. The annulus $\A(R)=\{u\in \C\,|\,
\frac{1}{R}<|u|<R\}$ of modulus $\mu=2\log R>0$ admits a
unique complete hyperbolic metric, which is given by \cite{BM}:
\ben
\label{gA}
\dd s_R^2=\lambda_R(u,\bar{u})^2|\dd u|^2~~,~\mathrm{where}~
~\lambda_R(u,\bar{u})=\frac{\pi}{2\log R}\frac{1}{|u|\cos\left(\frac{\pi\log|u|}{2\log R}\right)}~~.
\een
In particular, $\Re u$ and $\Im u$ are isothermal coordinates. Notice
that the transformation $u\rightarrow \frac{1}{\bar{u}}$ is an
isometry. In polar coordinates $(\rho,\theta)$ defined through:
\be
u=\rho e^{\i\theta}~,~\mathrm{with}~~\rho=|u|\in (\frac{1}{R},R)~~,
\ee
the metric takes the form: 
\ben
\dd s_R^2=\left(\frac{\pi}{2\log R}\right)^2\frac{\dd\rho^2+\rho^2\dd \theta^2}{\left[\rho \cos\left(\frac{\pi\log\rho}{2\log R}\right)\right]^2}~~.
\een
For any $\rho\in (1/R,R)$, the Euclidean circle $c_\rho\eqdef \{u\in
  \C||u|=\rho\}$ has hyperbolic circumference given by:
\be
\ell_R(c_\rho)=\frac{\pi^2}{(\log R) \cos\left(\frac{\pi\log \rho}{2\log R}\right)}~~.
\ee
We have:
\be
\ell_R(c_\rho)=\ell_R(c_{1/\rho})~~.
\ee
Notice that $\ell_R(c_\rho)$ increases from $\frac{\pi^2}{\log
  R}$ to infinity as $\rho$ increases from $1$ to $R$ and as $\rho$
decreases from $1$ to $1/R$. In particular, the minimum hyperbolic
length is attained for the circle $c_1$ of Euclidean radius $\rho=1$,
which is the only closed hyperbolic geodesic of $\A(R)$ and has
hyperbolic length:
\ben
\label{ellRC1}
\ell\eqdef \ell_R(c_1)=\frac{\pi^2}{\log R}=\frac{2\pi^2}{\mu}~~,
\een
known as the {\em hyperbolic circumference} of $\A(R)$. The lines:
\be
\{t e^{i\theta}|t\in (1/R,R)\}
\ee
are geodesics of infinite length. Notice that the hyperbolic area of
$\A(R)$ is infinite. Relation \eqref{ellRC1} gives: 
\ben
\label{Rell}
R=R_\ell=e^{\frac{~\pi^2}{\ell}}~~.
\een

\subsection{Diffeomorphism to the punctured disk and to the punctured plane}

\noindent
The annulus $\A(R)$ is diffeomorphic (but not biholomorphic !) with
the punctured unit disk with complex coordinate $u'$ through the map:
\ben
\label{anntopdisk}
u'=f(u)=\frac{|u|-\frac{1}{R}}{(R-\frac{1}{R})|u|}u\in \mD^\ast ~~(u\in \A(R))~~,
\een
whose inverse is given by: 
\be
u=f^{-1}(u')=\frac{(R-\frac{1}{R})|u'|+\frac{1}{R}}{|u'|} u'\in \A(R)~~(u'\in \mD^\ast)~~.
\ee
Notice that $f$ maps the Euclidean circle $|u|=\frac{1}{R}$ to $u'=0$
and the Euclidean circle $|u|=R$ to the Euclidean circle
$|u'|=1$. Composing $f$ with the map \eqref{rhomDast} gives a
diffeomorphism from $\A(R)$ to the punctured complex plane with
coordinate:
\be
\zeta=\fr e^{\i\theta}~~,
\ee
where: 
\ben
\label{frA}
\fr=-\frac{1}{\log|u'|}=\frac{1}{\log \frac{R-\frac{1}{R}}{|u|-\frac{1}{R}}}=\frac{1}{\log \frac{R-\frac{1}{R}}{\rho-\frac{1}{R}}}~~.
\een
The funnel end at $|u|=\frac{1}{R}$ corresponds to $\zeta=0$ while the
funnel end at $|u|=R$ corresponds to $\zeta=\infty$.

\subsection{The end compactification of $\A(R)$}

\noindent
The end compactification of $\A(R)$ is the unit sphere $\rS^2$ with
polar coordinates $(\psi,\theta)\in (0,\pi)\times (0,2\pi)$, which
maps to the $\zeta$-plane through the stereographic projection \eqref{sterproj}.
The funnel end at $|u|=\frac{1}{R}$ corresponds to the south pole, while 
the funnel end at $|u|=R$ corresponds to the north pole. The explicit 
embedding of $\A(R)$ into $\rS^2$ is given by: 
\be
\psi=2\arctan\left(\log \frac{R-\frac{1}{R}}{\rho-\frac{1}{R}}\right)~~,~~\theta=\theta~~.
\ee

\subsection{The hyperbolic funnel}

\noindent
Let $\ell>0$ be a positive real number and $R_\ell$ be defined as in
\eqref{Rell}. By definition, a {\em hyperbolic funnel}
(cf. \cite{genalpha}) of circumference $\ell>0$ is the annulus:
\ben
\label{F}
\rF_\ell=\{u\in \C\,\, |\,\, \frac{1}{R_\ell}<|u|<1\}\subset \A(R_\ell)~~,
\een
endowed with the restriction of the metric \eqref{gA}. Since
$u\rightarrow \frac{1}{\bar{u}}$ is an isometry of $\A(R)$, the funnel
is isometric with the annulus $1<|u|<R_\ell$ (endowed with the
restriction of \eqref{gA}). Hence $\A(R)$ decomposes as the disjoint union 
of two funnels and the closed geodesic $c_1$. Notice that a funnel has
infinite hyperbolic area. The funnel is diffeomorphic (but not
conformally equivalent !)  with the punctured unit disk through the
map:
\ben
\label{funneltopdisk}
u'=\frac{|u|-\frac{1}{R_\ell}}{(1-\frac{1}{R_\ell})|u|}u\in \mD^\ast ~~(u\in \rF_\ell)~~,
\een
which takes the funnel end $u\rightarrow \frac{1}{R_\ell}$ to the
center $u'\rightarrow 0$ of the unit disk and the circle $c_1$ into
the bounding circle of the unit disk.

\subsection{Semi-geodesic coordinates on $\rF_\ell$}

\noindent
Consider orthogonal coordinates $(q,\theta)$ on $\rF_\ell$ given by:
\be
q=\frac{\pi}{\mu}\frac{1}{\cos(\frac{\pi}{\mu}\log\rho)}\in (\frac{\pi}{\mu},+\infty)~~,
\ee
where $\rho=|u|$ and $\theta\in (0,2\pi)$ is the polar angle in the
$u$-plane. The limit $\rho\rightarrow 1/R$ corresponds to
$q\rightarrow \infty$, while $\rho\rightarrow 1$ corresponds to
$q\rightarrow \frac{\pi}{\mu}=\frac{\ell}{2\pi}$. In these
coordinates, the metric on $\rF_\ell$ becomes:
\ben
\label{fmetric2}
\dd s_{\rF_\ell}^2=\frac{1}{q^2-\left(\frac{\pi}{\mu}\right)^2}\dd q^2+q^2 \dd\theta^2=\frac{1}{q^2-\left(\frac{\ell}{2\pi}\right)^2}\dd q^2+q^2 \dd\theta^2~~.
\een
The coordinates $\theta$ and:
\ben
\label{rq}
r\eqdef \arccosh\left(\frac{2\pi}{\ell}q\right)=\arccosh\left[\frac{1}{\cos(\frac{\pi}{\mu}\log\rho)}\right]\in (0,+\infty)
\een
are semi-geodesic on $\rF_\ell$. In these
coordinates, the metric takes the form:
\ben
\label{fmetric1}
\dd s_{\rF_\ell}^2=\dd r^2+\frac{\ell^2}{(2\pi)^2}\cosh^2(r)\dd\theta^2~~.
\een
The limit $\rho\rightarrow 1/R$ corresponds to $r\rightarrow +\infty$
while $\rho\rightarrow 1$ corresponds to $r\rightarrow 0$. The
parameter $\ell>0$ is the hyperbolic length of the geodesic at $r=0$.

\subsection{Partial isometric embedding of the funnel into Euclidean 3-space}

\begin{figure}[H]
\centering \includegraphics[width=50mm]{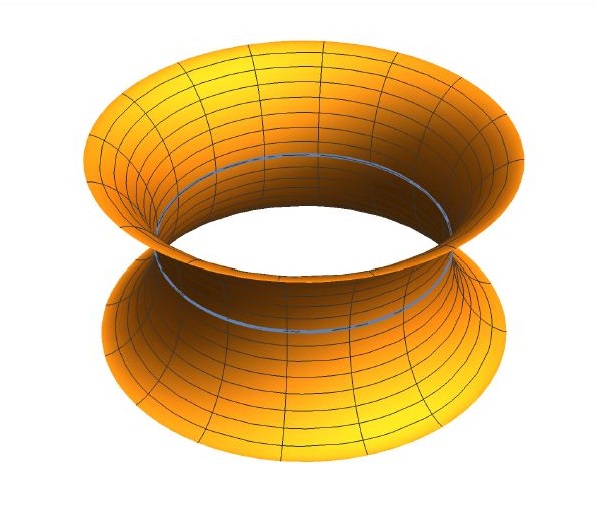}
\caption{Hyperbolic surface of revolution of hyperboloid type. The
  upper half corresponds to the region $\rF_\ell^0$ of $\rF_\ell$. The
  circle drawn in light blue is the closed geodesic $c_1$ of length $\ell$.}
\label{fig:hyprev}
\end{figure}

\noindent
One can isometrically embed the annulus $\rF_\ell^0\subset \rF_\ell$
corresponding to the range:
\be
\frac{\ell}{2\pi}=\frac{\pi}{\mu} < q\leq \sqrt{1+ \left(\frac{\pi}{\mu}\right)^2}=\sqrt{1+ \left(\frac{\ell}{2\pi}\right)^2}
\ee
into Euclidean $\R^3$ as the surface of revolution defined by the parametric
equations (see \cite[Chap. 3C]{Kuhnel} or \cite[Chap. 15]{Gray}):
\be
x_1=q(r)\cos \theta~~,~~x_2=q(r)\sin \theta~,~~x_3=\xi(r)~~,~\mathrm{with}~~r\in \left(0,\arcsinh\left(\frac{2\pi}{\ell}\right)\right)~~,
\ee
where:
\beqa
q(r)&=& \frac{\ell}{2\pi}\cosh(r)\in \left(\frac{\ell}{2\pi}, \sqrt{1+ \left(\frac{\ell}{2\pi}\right)^2}\right)~~(\mathrm{see}~~\eqref{rq})\nn\\
\xi(r) &\eqdef& \int_0^r\dd r' \sqrt{1-\left(\frac{\ell}{2\pi}\right)^2\sinh^2(r')}=-\i\, \mathrm{E}\left(\i r\, , -\left(\frac{\ell}{2\pi}\right)^2\right)~~,
\eeqa
and $\mathrm{E}(\tau,m)$ denotes the elliptic integral of the second
kind. This is one of the three\footnote{The other two are the
  pseudosphere/tractricoid (which corresponds to a portion of the
  hyperbolic cusp) and the surface of ``conical type''.}  types of
(incomplete) classical surfaces of revolution in $\R^3$ of constant
Gaussian curvature equal to $-1$, namely a surface of ``hyperboloid
type'' (see Figure \ref{fig:hyprev}).

\subsection{Canonical uniformization to $\H$}

\noindent
The hyperbolic annulus $\A(R_\ell)$ is uniformized to the upper half
plane by the hyperbolic cyclic subgroup generated by the
transformation:
\ben
\label{hypcan}
\tau \rightarrow e^\ell \tau~~,
\een
which corresponds to the hyperbolic element: 
\ben
\label{Hell}
H_\ell\eqdef \left[\begin{array}{cc} e^{\ell/2} & 0\\ 0 & e^{-\ell/2} \end{array}\right]\in \PSL(2,\R)~~
\een
and fixes the points $\tau=0$ and $\tau=\infty$ lying on
$\partial_\infty \H$. The uniformization map is:
\ben
\label{piannuluscan}
u=\pi_\H(\tau)=R_\ell e^{\frac{2\pi i}{\ell}\log \tau}=e^{\frac{\pi^2}{\ell}+\frac{2\pi i}{\ell}\log \tau}~~,
\een
which gives: 
\ben
\label{piAnnulus}
\rho=e^{\frac{\pi^2}{\ell}-\frac{2\pi}{\ell}\arg(\tau)}~~,~~\theta=\frac{2\pi}{\ell}\log|\tau|~~(\mathrm{mod}~2\pi)~~.
\een
A fundamental polygon is given by (see Figure \ref{fig:ADomain1}): 
\ben
\label{Dannulus}
\fD_\H = \left\{\tau\in\mathbb{H} \,\, |\,\, e^{l} < |\tau| < e^{2 l}\right\}~~, ~~l>0~~.
\een
This is a hyperbolic quadrilateral with two free sides and 
vertices located at the points: 
\ben
\label{ABCDAnnulus}
A:\{\tau=-e^{2\ell}\}~,~B:\{\tau=-e^{\ell}\}~,~C:\{\tau=e^{\ell}\}~~,~~D:\{\tau=e^{2\ell}\}~~.
\een
The funnel $\rF_\ell$ is the projection of the relative funnel
neighborhood \cite{genalpha}:
\be
\fF_\H^\ell=\{\tau \in \fD_\H|\Re \tau< 0\}=\{\tau\in \fD_\H | \arg (\tau)\in \left(\frac{\pi}{2}, \pi\right)\}~~.
\ee

\subsection{Canonical uniformization to $\mD$}

\begin{figure}[H]
\centering
\begin{minipage}{.47\textwidth}
\centering
\vskip 4mm
\includegraphics[width=.9\linewidth]{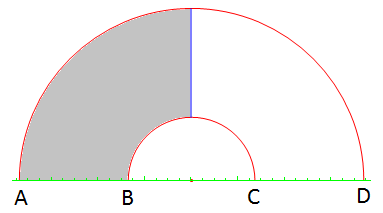}
\vskip 2em \subcaption{Fundamental domain $\fD_\H$ for $\A(R)$ on
  the Poincar\'e half-plane. The shaded region is the relative funnel
  neighborhood $\fF_\H$ corresponding to the free side on the left.}
\label{fig:ADomain1}
\end{minipage}
\hfill
\begin{minipage}{.47\textwidth}
\centering
\includegraphics[width=0.7\linewidth]{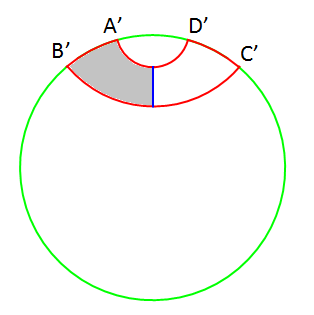}
\subcaption{Fundamental domain $\fD_\mD$ for $\A(R)$ on the Poincar\'e
  disk. The shaded region is the relative funnel neighborhood of the
  free side on the left.}
\label{fig:ADom2}
\end{minipage}
\caption{Fundamental domains for the uniformization of the hyperbolic
  annulus to the Poincar\'e half-plane and to the hyperbolic disk.}
\end{figure}

\noindent
When passing to the disk model, the fundamental domain $\fD_\H$ is
mapped by the Cayley transformation \eqref{Cayley} to a hyperbolic
quadrilateral $\fD_\mD$ with vertices located at the following points,
which are obtained from the points $A,B,C,D$ defined in
\eqref{ABCDAnnulus} by applying \eqref{Cayley}:
\be
A':\{u=-\frac{\i + e^{2l}}{\i e^{2l}+1}\}~~,~~B':\{u=-\frac{\i + e^{l}}{\i e^{l}+1}\}~~,~~C':\{u=\frac{\i - e^{l}}{\i e^{l}-1}\}~~,~~D':u=\{\frac{\i - e^{2l}}{\i e^{2l}-1}\}~~.
\ee
The free sides $(A'B')$ and $(C'D')$ are portions of
$\partial_\infty \mD$ (see Figure \ref{fig:ADom2}), while
the sides $(B'C')$ and $(A'D')$ are arc segments of Euclidean 
circles which are orthogonal to $\partial_\infty \mD$. 

\subsection{Globally well-behaved scalar potentials on $\A(R)$}

\begin{figure}[H]
\centering
\vskip -0.9em
\begin{minipage}{.47\textwidth}
\vskip 2em \centering \includegraphics[width=1 \linewidth]{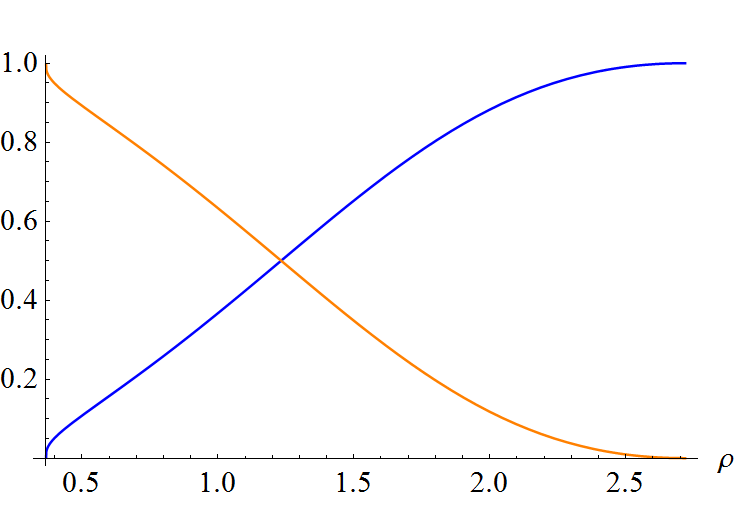}
\vskip 3em \subcaption{Plot of $\Phi_+/M_0$ (blue) and $\Phi_-/M_0$
  (yellow) as functions of $\rho\in (\frac{1}{R},R)=(e^{-1},e)$ for
  $\ell=\pi^2$. The values $\rho=\frac{1}{R}$ and $\rho=R$ correspond
  to the two funnel ends of $\A(R)$.}
\label{fig:PhipmAnnulus}
\end{minipage}\hfill
\begin{minipage}{.47\textwidth}
\centering \includegraphics[width=.95\linewidth]{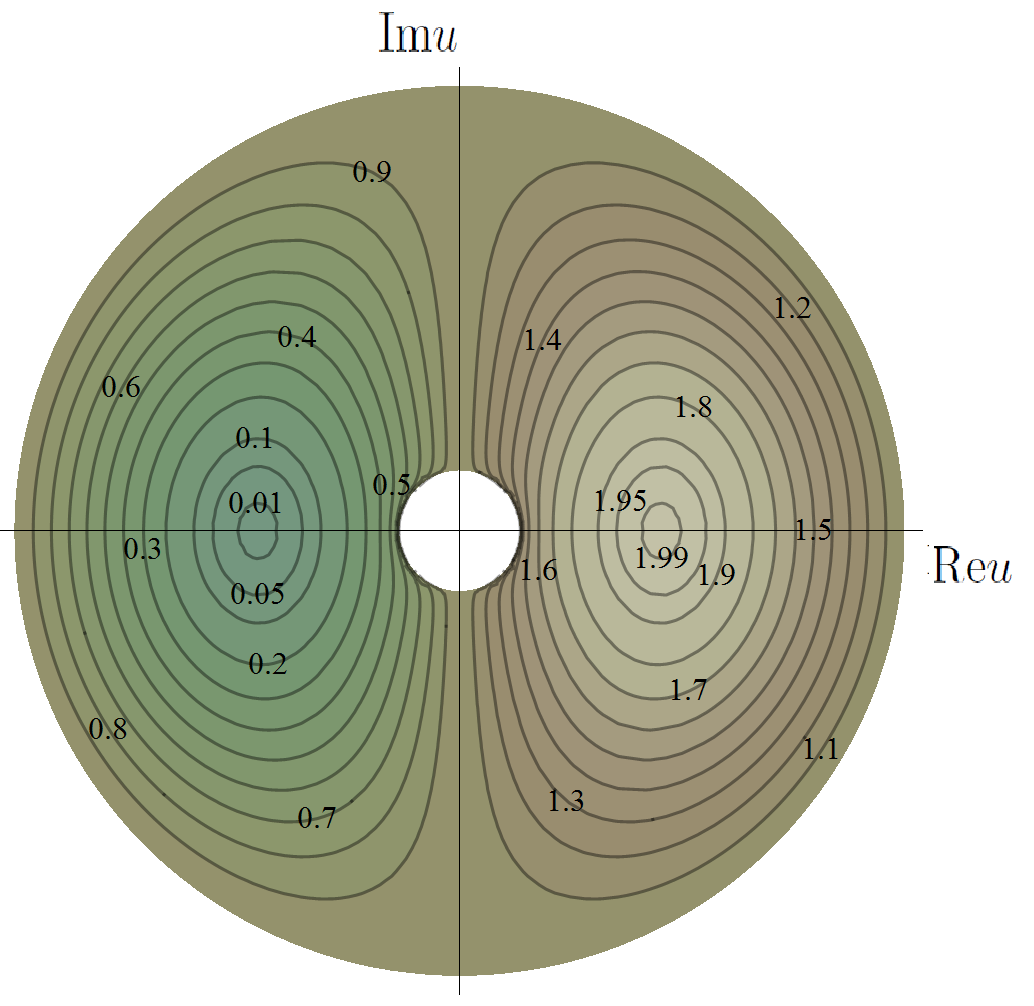}
\subcaption{Level plot of $\Phi_0/M_0$ on $\A(R)$ for
  $\ell=\pi^2$. Darker tones indicate lower values of $\Phi_0$. In
  this case, the maximum of $\Phi_0$ is located at
  $u_M=\frac{e^2+e-1}{e^2}$, while the minimum is located at
  $u_m=-u_M=-\frac{e^2+e-1}{e^2}$.}
\label{fig:Phi0Annulus}
\end{minipage}
\caption{The potentials $\Phi_\pm$ and $\Phi_0$ on the annulus of
  modulus $\mu=2$ ($\ell=\pi^2$, $R=e$). For $\Phi_+$ and $\Phi_-$, we
  indicate only the dependence of $\rho=|u|$, since these two
  potentials do not depend on the polar angle $\theta$. Notice that
  $\Phi_+$ tends to its infimum at the inner funnel end, while
  $\Phi_-$ tends to its infimum at the outer funnel end. Similarly,
  $\Phi_+$ and $\Phi_-$ tend to their suprema at opposite funnel
  ends. }
\end{figure}

\noindent
A scalar potential $\Phi$ on $\A(R)$ is globally well-behaved iff
there exists a smooth function $\hPhi:\rS^2\rightarrow \R$ such that:
\be
\Phi(\fr,\theta)=\hPhi(2\arccot(\fr),\theta)~~,
\ee
i.e.: 
\be
\Phi(\rho,\theta)=\hPhi\left(2\arctan\left(\log \frac{R-\frac{1}{R}}{\rho-\frac{1}{R}}\right),\theta\right)~~.
\ee
Expansion \eqref{hPhiExp} gives the
uniformly-convergent series:
\be
\Phi(\rho,\theta)=\sum_{l=0}^\infty \sum_{m=-l}^l D_{lm} Y_{lm}\left(2\arctan\left(\log \frac{R-\frac{1}{R}}{\rho-\frac{1}{R}}\right),\theta\right)~~.
\ee
For the choices \eqref{hPhipm}, we find: 
\ben
\label{APhipm}
\Phi_+=M_0\frac{\fr^2}{1+\fr^2}=M_0\frac{1}{1+ \left[\log \frac{R-\frac{1}{R}}{\rho-\frac{1}{R}}\right]^2}~~,
~~\Phi_-=M_0\frac{1}{1+\fr^2}=M_0\frac{\left[\log \frac{R-\frac{1}{R}}{\rho-\frac{1}{R}}\right]^2}{1+\left[\log \frac{R-\frac{1}{R}}{\rho-\frac{1}{R}}\right]^2}~~,
\een
where $\rho=|u|\in (\frac{1}{R},R)$ and we used \eqref{frA}. The
potentials $\Phi_\pm$ are plotted in Figure \ref{fig:PhipmAnnulus} for
the case $\ell=\pi^2$ ($R=e$). For the choice \eqref{hPhi0}, we find:
\ben
\label{APhi0}
\Phi_0=M_0\left[1+\frac{2\fr}{1+\fr^2}\cos\theta\right]=M_0\left[1+\frac{2\log\frac{R-\frac{1}{R}}{\rho-\frac{1}{R}}}{1+\left(\log\frac{R-\frac{1}{R}}{\rho-\frac{1}{R}}\right)^2}\cos\theta \right]~~.
\een
Recall that $\hPhi_0$ has two extrema on $\rS^2$, which are located at
$(\psi,\theta)=(\frac{\pi}{2},0)$ (maximum) and
$(\psi,\theta)=(\frac{\pi}{2},\pi)$ (minimum). At each of these
points, relation \eqref{sterproj} gives $\fr=\cot(\frac{\pi}{4})=1$,
so \eqref{frA} gives $\rho=\rho_0$, where:
\ben
\label{rho0}
\rho_0\eqdef \frac{1}{R}+\frac{1}{e}(R-\frac{1}{R})~~. 
\een
It follows that the two critical points of $\Phi_0$ on $\A(R)$ are
located on the real axis at:
\begin{itemize}
\itemsep 0.0em
\item $u_M=+\rho_0$, where $\Phi_0$ attains its maximum (which equals
  $2M$)
\item $u_m=-\rho_0$, where $\Phi_0$ attains its minimum (which equals
  zero).
\end{itemize}
The level curves of $\Phi_0$ are shown in Figure \ref{fig:Phi0Annulus}
for the case $\ell=\pi^2$ ($R=e$), which gives
$\rho_0=\frac{e^2+e-1}{e^2}=1+\frac{1}{e}-\frac{1}{e^2}\approx 1.23 $.

\

\

\paragraph{Lift of $\Phi_\pm$ and $\Phi_0$ to $\H$.}

\begin{figure}[H]
\centering
\vskip -.8em
\begin{minipage}{.45\textwidth}
\vskip 2em
\centering \includegraphics[width=1\linewidth]{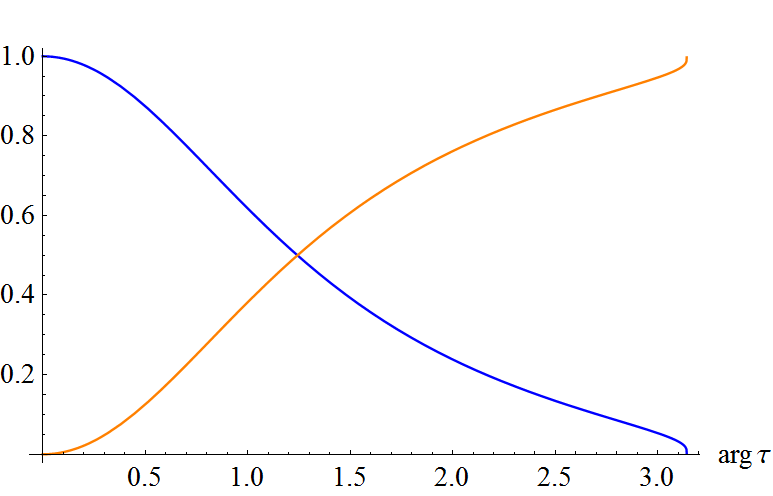}
\vskip 4em \subcaption{Plot of $\tPhi_+/M_0$ (blue) and
  $\tPhi_-/M_0$ (yellow) as functions of $\arg(\tau)\in (0,\pi)$ for
  $\ell=\pi^2$. The values $\arg(\tau)=0$ and $\arg(\tau)=\pi$
  correspond to the two funnel ends of $\A(R)$.}
\label{fig:tPhipmAnnulus}
\end{minipage}\hfill
\begin{minipage}{.5\textwidth}
\centering
\includegraphics[width=0.9\linewidth]{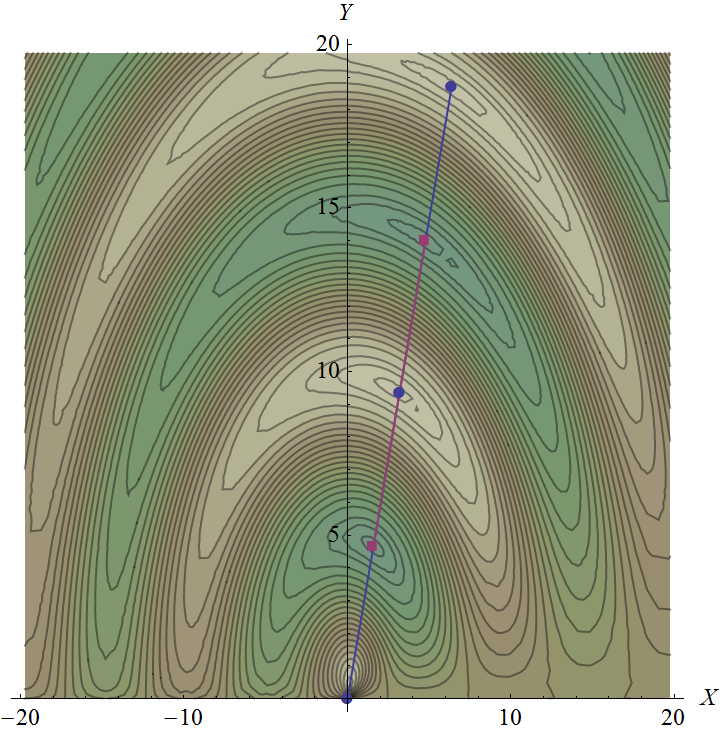}
\vskip -0.2em\subcaption{Level plot of $\tPhi_0/M_0$ in the coordinates $X,Y$ for
  $\ell=\pi^2$. Darker tones indicate lower values of $\tPhi_0$. The
  maxima of $\tPhi_0$ are indicated as blue dots, while the minima are
  shown as small red squares.}
\label{fig:tPhi0Annulus}
\end{minipage}
\vskip -0.7em
\caption{The lifted potentials $\tPhi_\pm$ and $\tPhi_0$ for the
  annulus of modulus $\mu=2$ ($\ell=\pi^2$, $R=e$).}
\end{figure}

The globally well-behaved scalar potentials \eqref{APhipm} and
\eqref{APhi0} lift to the following potentials on $\H$ (see Figures
\ref{fig:tPhipmAnnulus} and \ref{fig:tPhi0Annulus}):
\beqan
&& \tPhi_+(\tau)=M_0\frac{1}{1+ \left[\log \frac{R-\frac{1}{R}}{\rho(\tau)-\frac{1}{R}}\right]^2}~,
~\tPhi_-(\tau)=M_0\frac{\left[\log \frac{R-\frac{1}{R}}{\rho(\tau)-\frac{1}{R}}\right]^2}{1+\left[\log \frac{R-\frac{1}{R}}{\rho(\tau)-\frac{1}{R}}\right]^2}\nn\\
&&\tPhi_0(\tau)=M_0\left[1+\frac{2\log\frac{R-\frac{1}{R}}{\rho(\tau)-\frac{1}{R}}}{1+\left(\log\frac{R-\frac{1}{R}}{\rho(\tau)-\frac{1}{R}}\right)^2}\cos\left(
\frac{2\pi}{\ell}\log|\tau|\right) \right]~~,
\eeqan
where $\rho(\tau)=e^{\frac{\pi^2}{\ell}-\frac{2\pi}{\ell}\arg(\tau)}$
(see \eqref{piAnnulus}). The lifted potential $\tPhi_0$ has local
maxima at the inverse image points of $u_M=+\rho_0$ and local minima
at the inverse image points of $u_m=-\rho_0$ (see \eqref{rho0}). Using
\eqref{piAnnulus}, we find that these are located at:
\begin{itemize}
\itemsep 0.0em
\item (maxima) $\tau=\tau_M(n)\eqdef e^{n\ell+\i \frac{\ell}{2\pi}\log
  \left(\frac{R}{\rho_0}\right)}$ (where $\tPhi_0$ equals $2M_0$)
\item (minima) $\tau=\tau_m(n)\eqdef e^{(n+\frac{1}{2})\ell+\i
  \frac{\ell}{2\pi}\log \left(\frac{R}{\rho_0}\right)}$ (where
  $\tPhi_0$ vanishes)~~,
\end{itemize}
with $n\in \Z$ an arbitrary integer. Hence all extrema of $\tPhi_0$
lie on the half-line $L_0$ through the origin which makes an angle
$\theta_0=\frac{\ell}{2\pi}\log \left(\frac{R}{\rho_0}\right)$ with
the $x$ axis of the $\tau$-plane.  The maxima and minima alternate
along this half-line and the ratio between the absolute values of two
consecutive maxima or two consecutive minima equals $e^\ell$; in
particular, the extrema accumulate toward the origin along $L_0$. The
fundamental domain $\fD_\H$ contains exactly one of the minima, namely
that located at $\tau=e^{\frac{3}{2}\ell+\i \theta_0}$. On the other
hand, the two non-free sides of $\fD_\H$ contain the two consecutive
maxima located at $\tau=e^{\ell+\i \theta_0}$ and $\tau=e^{2\ell+\i
  \theta_0}$; these two maxima of $\tPhi_0$ are identified by the
projection $\pi_\H$.

In Figure \ref{fig:tPhi0Annulus}, we show the level curves of
$\tPhi_0$ for the case $\ell=\pi^2$ ($R=e$), which gives
$\rho_0=1+\frac{1}{e}-\frac{1}{e^2}\approx 1.23$ and $\theta_0=
\frac{\pi}{2}\left[1-\log\left(1+\frac{1}{e}-\frac{1}{e^2}\right)\right]\approx
0.39\pi$. In this case, the absolute value of the ratio of successive
minima or maxima of $\tPhi_0$ equals $e^{\pi^2}\approx 1.93 \times
10^4$. Due to the large size of this ratio, we chose for clarity to
display the level curves of $\tPhi$ on a region of the
``semi-logarithmic upper half plane''. The latter has complex
coordinate $T=X+iY$ (where $X=\Re T$ and $Y=\Im T>0$), being related
to the region:
\be
\cR\eqdef \{\tau\in \H\,|\,|\tau|>1\}
\ee
of the Poincar\'e half-plane through the coordinate
transformation:
\be
T=(\log|\tau|)e^{\i\arg(\tau)}~~,~~\tau=e^{|T|} e^{\i\arg(T)}~~.
\ee
Notice that $\cR$ contains only those extrema of $\tPhi_0$ which have
absolute value larger than one. In the semi-logarithmic half-plane,
these extrema are located at $T=n\ell e^{\i \theta_0}$ with $n\in
\Z_{>0}$ (maxima) and $T=(n+\frac{1}{2})\ell e^{\i \theta_0}$ with
$n\in \Z_{\geq 0}$ (minima), lying equally-spaced on the half-line in
the $T$-plane which passes through the origin at angle $\theta_0$ with
the $X$ axis. The Cartesian coordinates $X,Y$ and $x=\Re\tau$, $y=\Im
\tau$ are related through:
\be
x=X \frac{e^{\sqrt{X^2+Y^2}}}{\sqrt{X^2+Y^2}}~~,~~y=Y \frac{e^{\sqrt{X^2+Y^2}}}{\sqrt{X^2+Y^2}}~~.
\ee
Figure \ref{fig:tPhi0Annulus} shows the level curves of the function
$\tPhi_0(X,Y)$ in the region defined by $|X|<2\ell$ and $|Y|<2\ell$
(where $2\ell=2\pi^2\approx 19.7$), which contains the image of the
following annular region of the Poincar\'e half-plane:
\be
A\eqdef \{\tau\in \H| 1<|\tau|<e^{2\ell}\}~~.
\ee
Notice that $A$ contains a copy of the fundamental domain $\fD_\H$.

\subsection{Lift of the cosmological model to $\H$}
\label{subsec:TrajAnnulus}

\noindent We now present examples of trajectories on $\A(R)$ for $\ell=\pi^2$
($R=e$, modulus $\mu=2$) for the vanishing scalar potential and for
the globally well-behaved scalar potentials $\Phi_\pm$ and
$\Phi_0$. These were obtained as explained in Subsection
\ref{subsec:lift}, by numerically computing solutions of the system
\eqref{elplane0} on the Poincar\'e half-plane for the corresponding
lifted potentials and then projecting these trajectories to the
hyperbolic punctured disk using the explicitly-known uniformization
map \eqref{piannuluscan}, which is equivalent with \eqref{piAnnulus}.

\paragraph{Trajectories for vanishing scalar potential.}

\begin{figure}[H]
\centering
\begin{minipage}{.47\textwidth}
\centering \includegraphics[width=.9\linewidth]{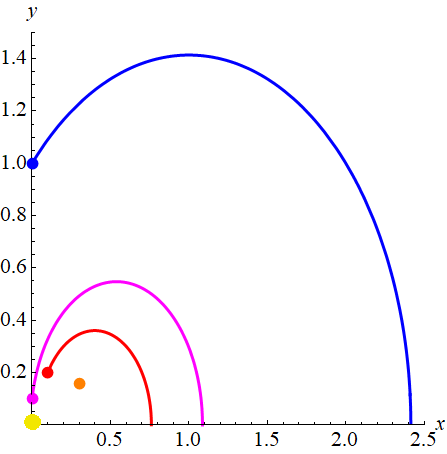}
\subcaption{For $\Phi=0$, the five trajectories on $\H$ with initial
  conditions given in Table \ref{table:InCond} coincide with those
  shown in Figure \ref{fig:PuncDiskNoPot}. }
\label{fig:AnnulusNoPot}
\end{minipage}
\hfill
\begin{minipage}{.47\textwidth}
\centering\includegraphics[width=1\linewidth]{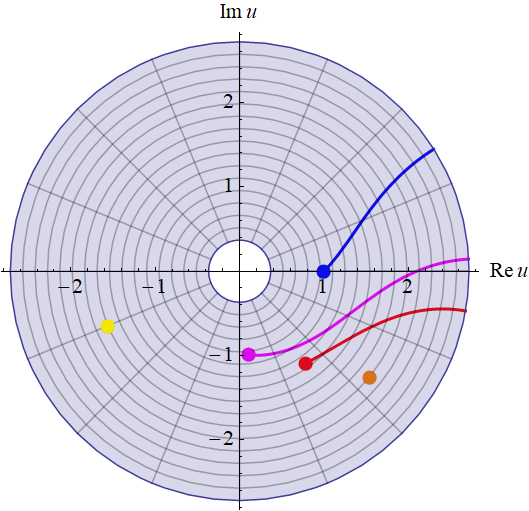}
\subcaption{Projection to $\A(R)$ of the trajectories shown at the
  left.  The projection differs from that for $\mD^\ast$ (cf. Figure
  \ref{fig:PuncDiskNoPotProj}) since the uniformization maps are
  different.}
\label{fig:AnnulusNoPotProj}
\end{minipage}
\caption{Five trajectories for $\Phi=0$, $\alpha=\frac{M_0}{3}$ and $\ell=\pi^2$, with
  initial conditions as in Table \ref{table:InCond}.}
\label{fig:AnnulusNoPotAll}
\end{figure}

Figure \ref{fig:AnnulusNoPotAll} shows five trajectories (orange, yellow, red,
blue and magenta) for $\alpha=\frac{M_0}{3}$, $\ell=\pi^2$ ($R=e$) and
$\Phi=0$, with the initial conditions given in Table
\ref{table:InCond} of Subsection \ref{subsec:TrajPuncDisk}. In this
case, vanishing initial velocity leads to two stationary trajectories (the
orange and yellow dots), while the hyperbolic geometry produces an effective
attraction toward the outer funnel end for $|u|>1$ and toward the
inner funnel end for $|u|<1$. The red, blue and magenta trajectories
start in the funnel region given by $|u|>1$ (with velocities pointing 
toward the outer funnel end) and hence evolve toward that end. 

\paragraph{Trajectories for $\Phi_-$.}

\begin{figure}[H]
\centering
\begin{minipage}{.45\textwidth}
\centering 
\includegraphics[width=.93\linewidth]{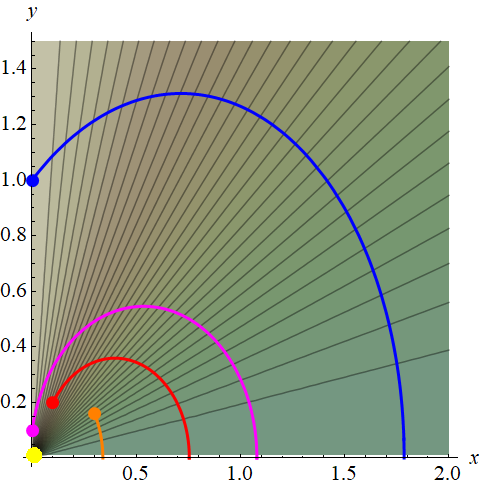}

\subcaption{Trajectories on $\H$ for $\tPhi=\tPhi_-$, superimposed on a 
level plot of $\tPhi_-$ on $\H$.}
\label{fig:AnnulusPhiMinus}
\end{minipage}
\hfill
\begin{minipage}{.47\textwidth}
\centering
\includegraphics[width=.93\linewidth]{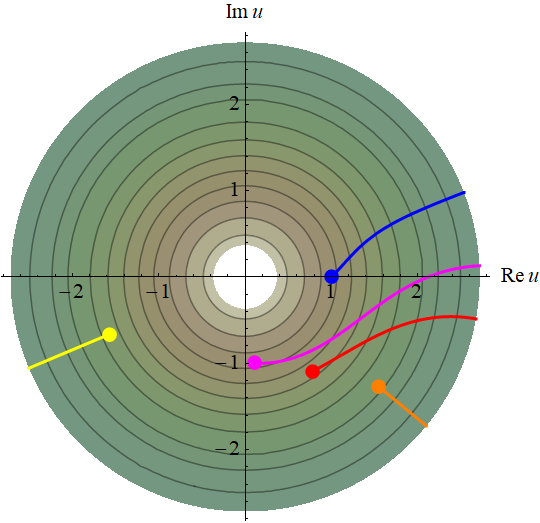}
\subcaption{Projection to $\A(R)$ of the trajectories shown at the left, 
superimposed on a 
level plot of $\Phi_-$ on $\A(R)$.}
\label{fig:AnnulusPhiMinusProj}
\end{minipage}
\caption{Five trajectories for $\Phi=\Phi_-$, $\alpha=\frac{M_0}{3}$ 
and $\ell=\pi^2$ with initial conditions of Table \ref{table:InCond}.}
\label{fig:AnnulusPhiMinusAll}
\end{figure}

Figure \ref{fig:AnnulusPhiMinusAll} shows five trajectories (orange, yellow
red, blue and magenta) for $\alpha=\frac{M_0}{3}$, $\ell=\pi^2$
($R=e$) and $\Phi=\Phi_-$, with the initial conditions given in Table
\ref{table:InCond}. In this case, the potential produces a repulsive
force away from the inner funnel end and an attraction force toward
the outer end, thus accentuating the effect of the hyperbolic metric 
on the trajectories shown.

\begin{figure}[H]
\centering

\begin{minipage}{.45\textwidth}
\centering 
\includegraphics[width=.93\linewidth]{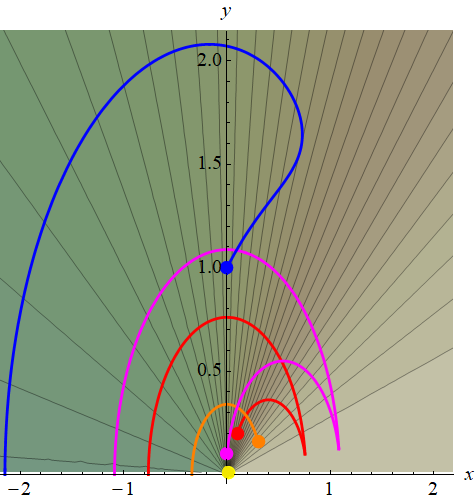}
\vskip 2mm
\subcaption{Trajectories on $\H$ for $\tPhi=\tPhi_+$, 
superimposed on a level plot of $\tPhi_-$ on $\H$. }
\label{fig:AnnulusPhiPlus}
\end{minipage}
\hfill
\begin{minipage}{.47\textwidth}
\centering
\includegraphics[width=.93\linewidth]{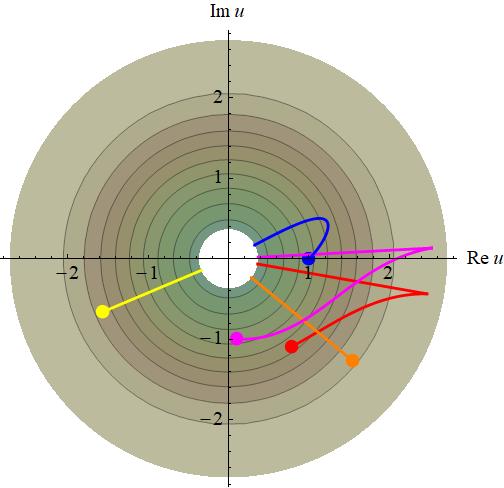}
\vskip 2mm
\subcaption{Projection to $\A(R)$ of the trajectories shown at the left, 
superimposed on a level plot of $\Phi_+$ on $\A(R)$.}
\label{fig:AnnulusPhiPlusProj}
\end{minipage}

\caption{Five trajectories for $\Phi=\Phi_+$, $\alpha=\frac{M_0}{3}$ and 
$\ell=\pi^2$, with initial conditions of Table \ref{table:InCond}.}
\label{fig:AnnulusPhiPlusAll}
\end{figure}

\paragraph{Trajectories for $\Phi_+$.}

Figure \ref{fig:AnnulusPhiPlusAll} shows five trajectories (orange, yellow
red, blue and magenta) for $\alpha=\frac{M_0}{3}$, $\ell=\pi^2$
($R=e$) and $\Phi=\Phi_+$, with the initial conditions given in Table
\ref{table:InCond}. In this case, the potential induces an attractive
force toward the inner funnel end. As a consequence, each of the
red, blue and magenta trajectories turns at some point in $\A(R)$ 
and evolves back toward the inner funnel end.

\begin{figure}[H]
\centering
\begin{minipage}{.47\textwidth}
\centering \includegraphics[width=.9\linewidth]{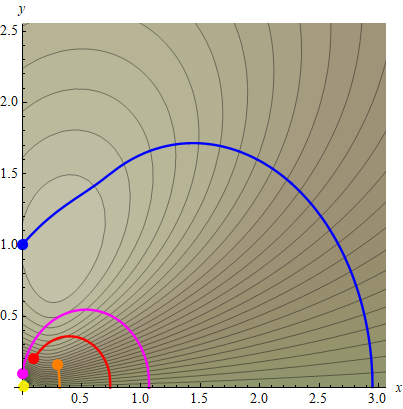}
\subcaption{Trajectories on $\H$ for $\tPhi=\tPhi_0$.
The yellow trajectory is not stationary, although at the scale shown it looks like a dot.  }
\label{fig:AnnulusPhi0}
\end{minipage}
\hfill
\begin{minipage}{.47\textwidth}
\centering \includegraphics[width=.9\linewidth]{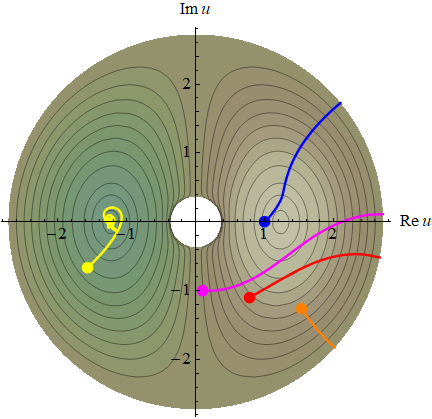}
\subcaption{Projection to $\A(R)$ of the five trajectories shown at
  the left. }
\label{fig:AnnulusPhi0Proj}
\end{minipage}
\caption{Five trajectories for $\Phi=\Phi_0$ when $\alpha=\frac{M_0}{3}$ and $\ell=\pi^2$.}
\label{fig:AnnulusPhi0All}
\end{figure}

\begin{figure}[H]
\centering
\begin{minipage}{.47\textwidth}
\centering \includegraphics[width=.85\linewidth]{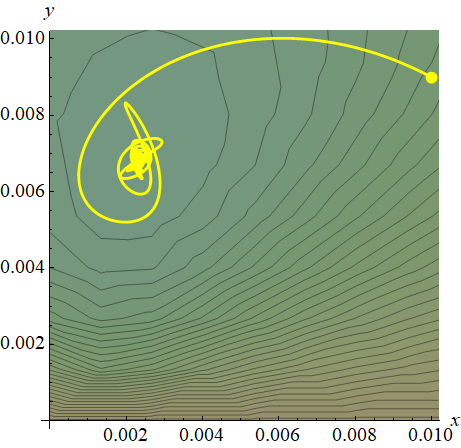}
\subcaption{Detail of the lifted yellow trajectory shown in Figure
  \ref{fig:AnnulusPhi0} near one minimum of $\tPhi_0$.} 
\vskip 12mm
\label{fig:TrajDetailAnnulus}
\end{minipage}\hfill
\begin{minipage}{.48\textwidth}
\centering \includegraphics[width=.86\linewidth]{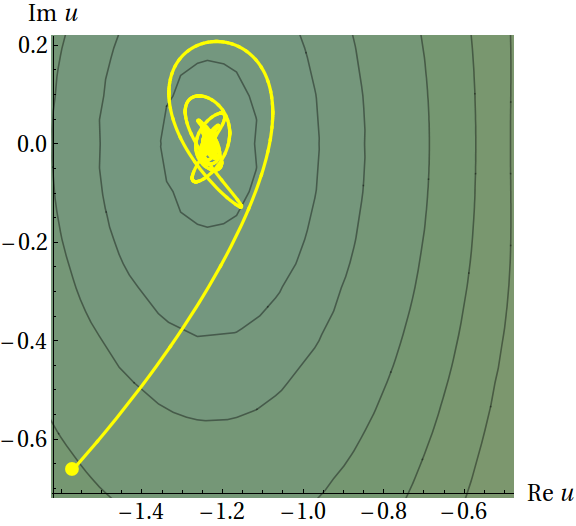}
\vskip 2.0mm \subcaption{Projection of the yellow trajectory to the hyperbolic annulus. 
The trajectory spirals in a
  complicated manner around the minimum of $\Phi_0$, until it settles
  at the minimum point.}
\vskip 3.0mm
\label{fig:TrajDetailAnnulusProj}
\end{minipage}
\caption{Detail of the yellow trajectory shown in Figure
  \ref{fig:AnnulusPhi0All}.}
\label{fig:TrajDetailAnnulusAll}
\end{figure}
\paragraph{Trajectories for $\Phi_0$.} 

Figure \ref{fig:AnnulusPhi0All} shows five trajectories (orange,
yellow, red, blue and magenta) for $\alpha=\frac{M_0}{3}$,
$\ell=\pi^2$ ($R=e$) and $\Phi=\Phi_0$, with the initial conditions
given in Table \ref{table:InCond}.  Figure
\ref{fig:TrajDetailAnnulusAll} shows a detail of the yellow trajectory
on both $\H$ and $\A(R)$.  It spirals in a complicated manner around
a minimum point of $\tPhi_0$ 
and projects to the single minimum of $\Phi_0$ on $\A(R)$ located
at $u=-\rho_0\approx -1.23$.  For $t\rightarrow \infty$, the
trajectory reaches the minimum point.

\subsection{Inflationary regions and the number of e-folds}
\label{subsec:epdisk}

\noindent Using relations \eqref{Hubble}-\eqref{efolds}, we find that
among the five trajectories with initial conditions given in Table
\ref{table:InCond}, the orange and yellow trajectories start in
inflationary regime for all three potentials $\Phi_+,\Phi_-$ and
$\Phi_0$, while the blue trajectory is always inflationary in
potential $\Phi_0$ but is never inflationary for the other
potentials. The orange and yellow trajectories give respectively $76$
and $74$ e-folds in the potential $\Phi_+$ after the first
inflationary regime\footnote{It is easy to see that one can rescale
the potential $\Phi_+$ by a positive constant in order to reach a
phenomenologically appropriate value $N\in [50,60]$ for each of the
orange and yellow trajectories in this potential.}, which lasts a
cosmological time of $t=2223 \,s $, respectively $t=2184 \, s$. The
other trajectories do not start in the inflationary regime for any of
the three potentials.

\begin{figure}[H]
\centering \includegraphics[width=60mm]{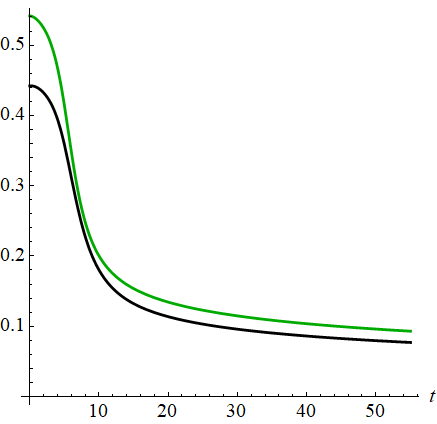}~\centering \includegraphics[width=60mm]{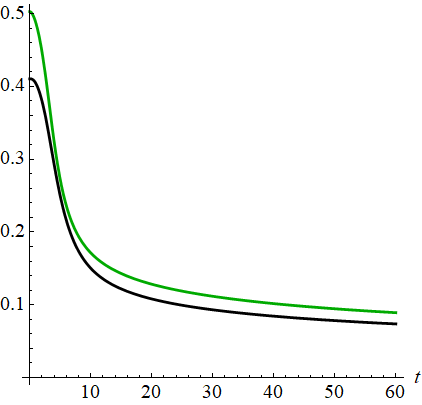}
\caption{Plots of $H(t)/\sqrt{M_0}$ (black) and $H_c(t)/\sqrt{M_0}$
(green) for the orange and yellow trajectories in the potential
$\Phi_+$. For better clarity, we truncated the plot at $t=60 \,s$. For
both trajectories, inflation stops after a finite time (indicated in
the text), though this is not visible at the scale of the figures.}
\label{fig:NHubPDphiP}
\end{figure}

\section{On the relation to observational cosmology}

\noindent As explained in \cite{genalpha}, generalized
$\alpha$-attractor models based on any geometrically-finite
non-compact hyperbolic Riemann surface $\Sigma$ and with a
well-behaved scalar potential enjoy universal behavior near each end
of $\Sigma$ where the extended potential has a local maximum, 
in a certain one-field truncation near that end. Within
this approximation, such models make the same predictions for the
spectral index $n_s$ and the tensor to scalar ratio $r$ as ordinary
$\alpha$-attractors, provided that inflation takes place sufficiently
close to such an end and along a trajectory which proceeds
radially from that end in canonical local semi-geodesic
coordinates. In the slow-roll approximation for motion along such a
trajectory, one finds \cite{genalpha}:
\ben
\label{obs}
n_s\approx 1-\frac{2}{N}~~,~~r\approx \frac{12\alpha}{N^2}~~,
\een
where $N$ is the number of e-folds during the inflationary period. For
such trajectories, generalized $\alpha$-attractors are therefore as
promising for matching observational data as ordinary
$\alpha$-attractors, whose agreement with current observations is
quite good \cite{alpha1,alpha2,alpha3,alpha4}. For the models
considered in this paper, such special trajectories are the radial
trajectories on the punctured disk and on the annulus, when inflation
takes place close to any of the components of the conformal
boundary. We gave two explicit examples of such trajectories:
\begin{itemize}
\item The yellow trajectory for the hyperbolic punctured disk in
potential $\Phi_+$ (see Figure \ref{fig:PuncDiskPhiPlusProj} in
Subsection 5.11). As discussed in Subsection \ref{subsec:edisk}, this
trajectory produces $55$ e-folds, a value which lies in the
observationally expected range of $50-60$ e-folds.
\item The orange and yellow trajectories for the hyperbolic annulus in
potential $\Phi_+$ (see figure \ref{fig:AnnulusPhiPlusProj} in
Subsection 6.10). As discussed in Subsection \ref{subsec:epdisk}, both
of these trajectories produce around $75$ e-folds, but a constant
positive rescaling of the potential $\Phi_+$ allows one to bring the
number of e-folds within the phenomenologically desired range $N\in
[50, 60]$.
\end{itemize}

Unlike the one-field $\alpha$-attractors usually considered in the
literature, generalized $\alpha$-attractors are genuine {\em
two-field} models and hence they can incorporate corrections to the
traditional paradigm of inflationary cosmology, which assumes for
simplicity that the inflaton is a single real scalar field. While
current observational data can be successfully reproduced by various
one-field models, they can also be reproduced by multi-field models
and it is generally deemed quite possible that, within the next
decade, improved measurements could detect deviations from one-field
model predictions. The recognition of this possibility has lead to
renewed interest in the study of multi-field models \cite{m2,m3,m4,m6}
and in particular to the development of numerical methods for
determining the effect of their cosmological perturbations
\cite{Dias1,Dias2,Mulryne} beyond the limitations of the SRST
approximation \cite{PT1,PT2}. As a single example, see \cite{c1} for a
recent investigation of constraints imposed on such models by Planck
2015 data \cite{Planck}.

In the context of the elementary generalized $\alpha$-attractor models
considered in this paper, deviations from the one-field paradigm could
be visible, for example, for trajectories which depart slightly from
the radial trajectories discussed above. This will affect the
two-point correlators which determine cosmological observables, thus
producing sub-leading corrections to relations
\eqref{obs}. Generalized $\alpha$-attractor models are also
interesting for investigations of the post-inflationary period of a
given trajectory, for which generic two field models have low
predictive power since they depend on the choice of an arbitrary
metric for the target manifold of the scalar fields. By contrast,
generalized $\alpha$-attractors provide a natural class of two-field
models which have universal behavior in the truncated inflationary
regime near the ends, while at the same time allowing for remarkable
dynamical complexity beyond that regime. As shown in the previous
sections, even the simplest instances of such models (namely those
based on elementary hyperbolic surfaces) already allow trajectories of
considerable complexity, due to the interplay between the effective
force induced by the hyperbolic geometry and that induced by the
scalar potential. Such models could therefore play the role of a
natural testing ground of two-field model technology, in a
mathematically tractable framework which may allow one to develop
insights deeper than those afforded by current approximations and by
generic numerical methods.

We end by mentioning that it is a non-trivial task to embed cosmological
models with a single real scalar field within fundamental quantum
theories of gravity such as string theory in a manner which is
compatible with all phenomenological and self-consistency
constraints. In particular, most scalar fields which arise naturally
in closed string theory are complex-valued and one has to rely on
special and quite finely tuned constructions when embedding single
field models in string theory in a consistent and phenomenologically
reasonable manner. Naturality arguments might therefore suggest that
the inflaton could in fact be a complex-valued field in a fundamental
theory of gravity, thus leading to a two-field cosmological model. In
this context, we mention that generalized $\alpha$-attractors appear
to have natural string-theoretic realizations which involve F-theory
backgrounds with discrete fluxes, though a proper discussion of that
construction (which involves the theory of modular curves and Shimura
varieties) lies well outside the scope of the present paper.

\section{Conclusions and further directions}

\noindent We studied generalized $\alpha$-attractor models defined by elementary
hyperbolic surfaces, showing how they fit into the framework developed
in \cite{genalpha}. Following a ``universal'' approach to globally
well-behaved scalar potentials, we showed how they can be approximated
systematically using the Laplace expansion of their extension to the
end compactification (which in such models is the unit sphere) and how
a smooth real-valued map defined on the latter induces different
potentials on each elementary hyperbolic surface. We also illustrated
cosmological dynamics of generalized $\alpha$-attractor models by
numerically extracting various trajectories for the cases of
$\mD^\ast$ and $\A(R)$, finding rather complex behavior even for
relatively simple scalar potentials. From the universal perspective
followed here and in \cite{genalpha}, the difference between models
with globally well-behaved scalar potential defined on various
hyperbolic surfaces of the same genus is captured by two maps, namely
the uniformization map $\pi_\H$ and the map $j$ which embeds the given
surface into its end compactification. For elementary surfaces, both
of these maps can be constructed explicitly and hence their effects on
the cosmological dynamics can be explored systematically.

A similar approach could in principle be followed for any
geometrically finite hyperbolic surface. In this regard, it would be
natural to explore the large class of non-elementary planar surfaces,
which form a classical subject in complex analysis and uniformization
theory. For such surfaces, the uniformization map $\pi_\H$ is not
usually known explicitly and, at least for general values of the
moduli, it must be determined numerically. Despite this fact, a
fundamental domain is known for any planar surface, as are certain
other properties of the hyperbolic metric and of the uniformization
map \cite{Hempel,SV}. This allows one to approach generalized
$\alpha$-attractor models whose scalar manifolds are given by such
surfaces using the general algorithm proposed in
\cite[Sec. 7]{genalpha}. For example, it would be interesting to
perform a detailed study of cosmological trajectories for some
triply-connected non-elementary planar surfaces such as the
twice-punctured disk \cite{Beardon2pdisk,HS1,HS2,HS3} and
once-punctured annulus \cite{Zhang}.

We also commented briefly on the potential relevance of such models to
observational cosmology. As pointed out in Section 7, the models
considered in this paper share the general features discussed in
\cite{genalpha} and hence provide reasonable candidates for
reproducing current observational constraints, similar to ordinary
$\alpha$-attractors. In particular, they easily support cosmological
trajectories which produce the expected number of $50-60$ e-folds. Of
course, much deeper investigation of such models is needed before the
question of their phenomenological relevance can be answered fully.

\acknowledgments{\noindent The work of M.B. and C.I.L. was supported by
  grant IBS-R003-S1. The authors thank C.~S.~Shahbazi for
  participation in the initial stages of the project.}

\end{document}